\documentclass[a4paper]{egpubl}
\usepackage{eurovis2019}

\SpecialIssuePaper         

\electronicVersion 
\PrintedOrElectronic

\usepackage{graphicx}
\DeclareGraphicsExtensions{.pdf,.png,.jpg,.jpeg}

\graphicspath{{figures/}{pictures/}{images/}{./}} 

\usepackage{url}
\usepackage[utf8]{inputenc}
\usepackage{microtype}                 
\PassOptionsToPackage{warn}{textcomp}  
\usepackage{textcomp}                  
\usepackage{mathptmx}                  
\usepackage{times}                     
\usepackage{cite}                      
\usepackage{booktabs}                  
\usepackage{cleveref}
\usepackage{caption}
\usepackage[labelfont=bf]{subcaption}
\usepackage{soul}
\usepackage{color}
\usepackage[dvipsnames]{xcolor}
\usepackage{t1enc,dfadobe}
\usepackage{egweblnk}
\usepackage{xspace}
\usepackage[frozencache]{minted}
\def\stampede{\emph{Stampede2}\xspace}
\def\theta{\emph{Theta}\xspace}

\def\RR{\textsuperscript\textregistered\xspace}
\def\TM{\textsuperscript\texttrademark\xspace}
\def\INTEL{Intel\RR}
\def\INTELXEON{\INTEL Xeon\RR}
\def\INTELXEONPHI{\INTEL Xeon Phi\TM}

\definecolor{dkgreen}{rgb}{0,0.6,0}
\definecolor{dkblue}{rgb}{0,0,0.6}
\definecolor{gray}{rgb}{0.5,0.5,0.5}
\definecolor{mauve}{rgb}{0.58,0,0.82}

\title[Scalable Ray Tracing Using the Distributed FrameBuffer]
{Scalable Ray Tracing Using the Distributed FrameBuffer\vspace{-1em}}

\author[Usher et al.]{
\vspace{-1em}
	Will Usher\thanks{will@sci.utah.edu}$^{1,2}$,
Ingo Wald$^{2,3}$,
Jefferson Amstutz$^2$,
Johannes G\"unther$^2$,
Carson Brownlee$^2$,
and Valerio Pascucci$^1$\\
$^1$SCI Institute, University of Utah \quad $^2$Intel Corporation \quad $^3$Now at NVIDIA}

\begin{document}


\teaser{
	\vspace{-3.5em}
	\centering
	\begin{subfigure}{0.4\textwidth}
		\centering
		\includegraphics[width=\textwidth,trim={0 35cm 0 4cm},clip]{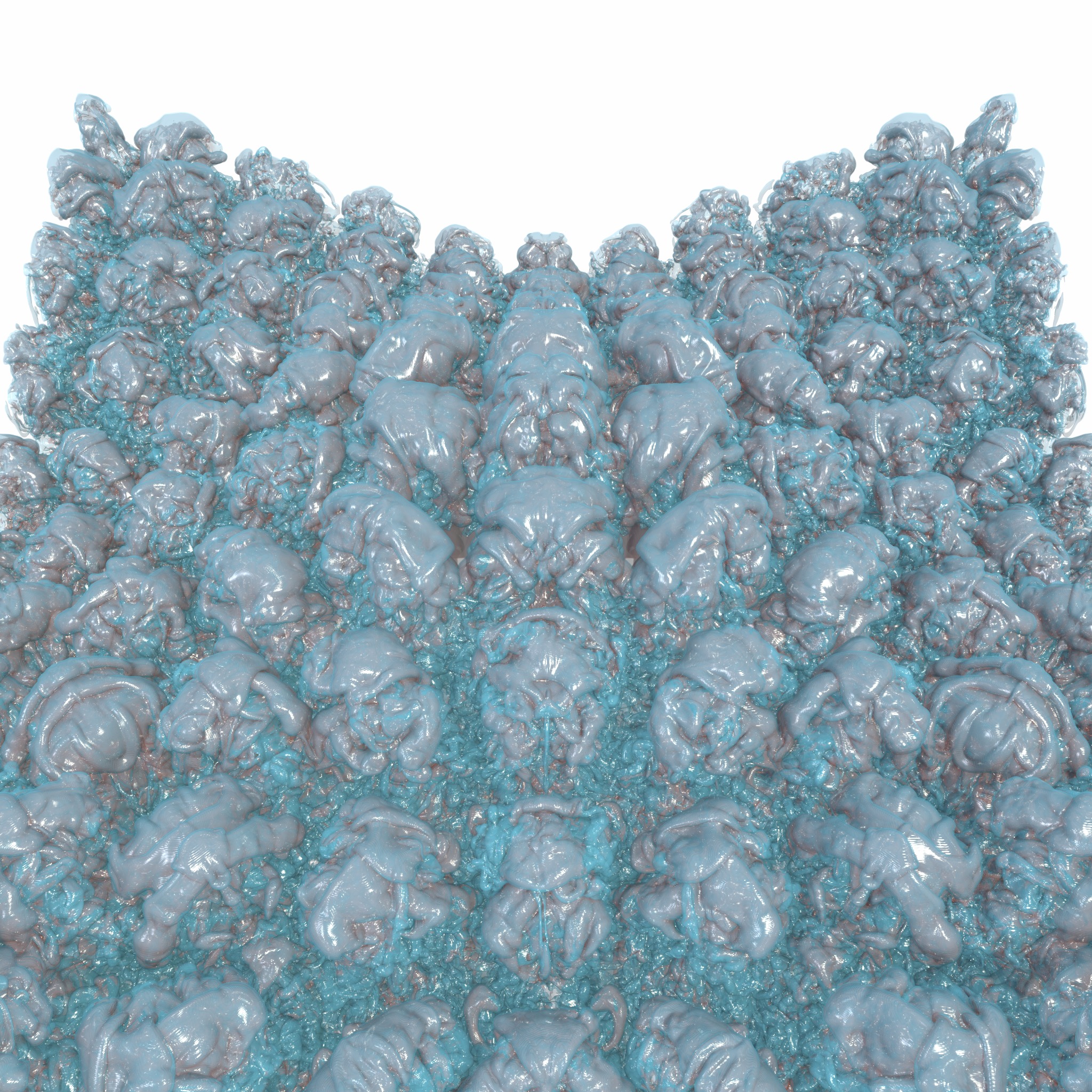}
	\end{subfigure}%
	\begin{subfigure}{0.4\textwidth}
		\centering
		\includegraphics[width=\textwidth,trim={0 35cm 0 4cm},clip]{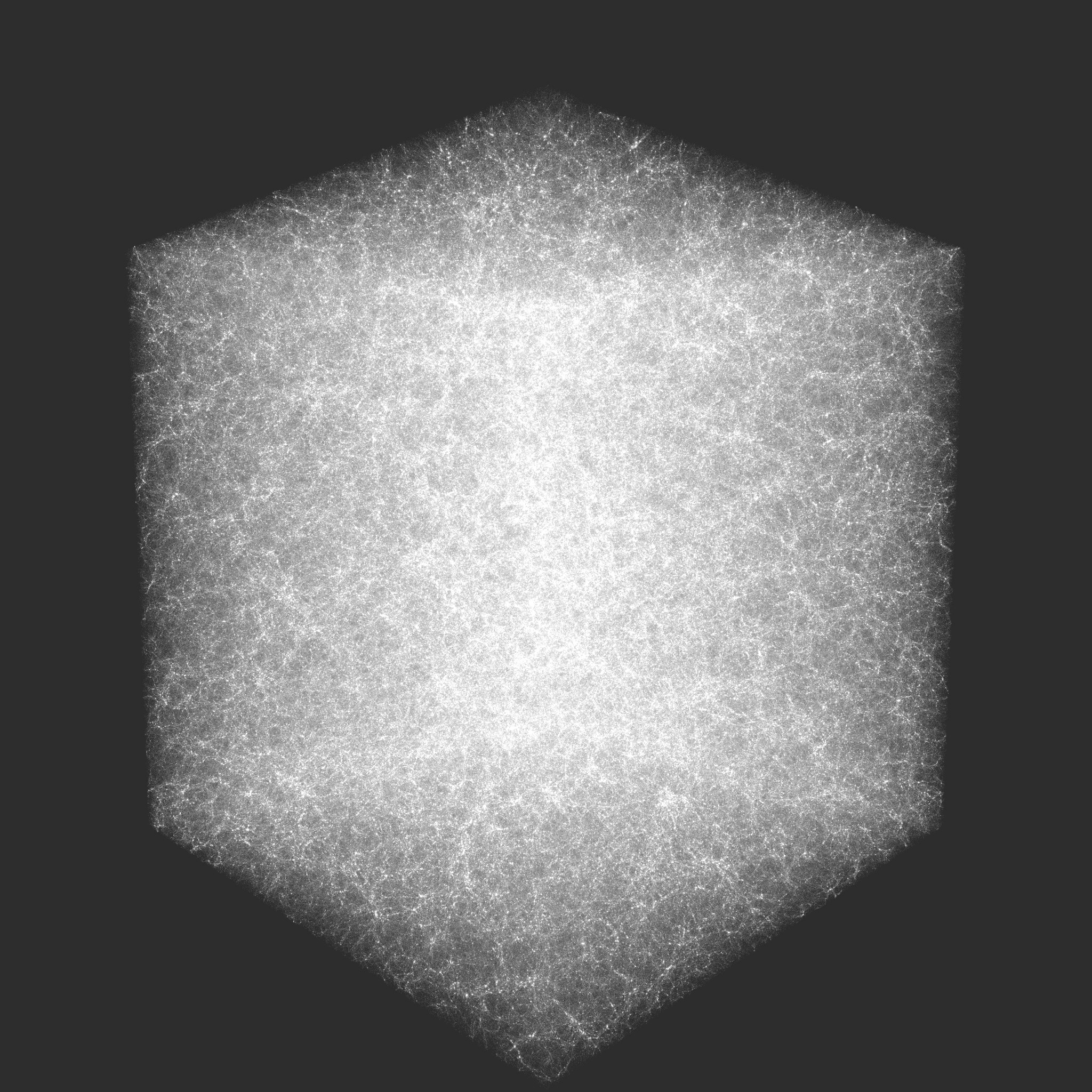}
	\end{subfigure}
	\begin{subfigure}{0.8\textwidth}
		\centering
		\includegraphics[width=\textwidth]{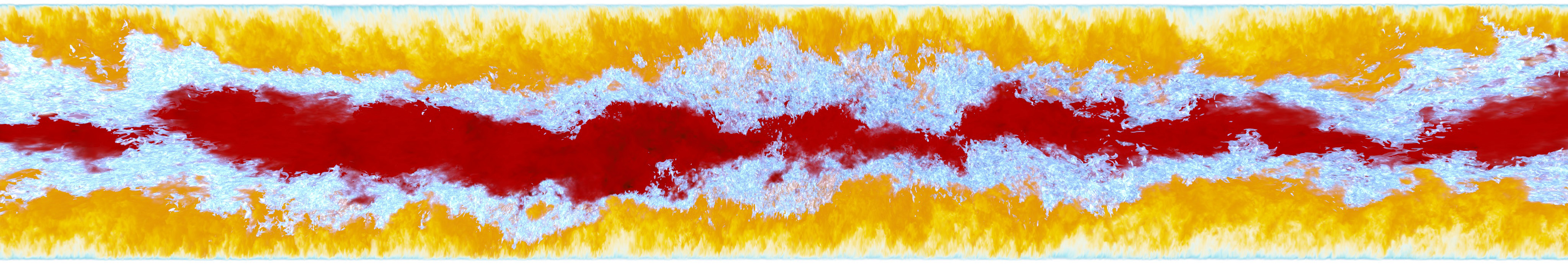}
	\end{subfigure}
	\vspace{-1em}
	\caption{\label{fig:teaser}%
	Large-scale interactive visualization using the Distributed FrameBuffer.
	Top left: Image-parallel rendering of two transparent isosurfaces from
	the Richtmyer-Meshkov~\cite{cohen_rm_2002} (516M triangles),
	8FPS with a $2048^2$ framebuffer using 16 \stampede \INTELXEON Platinum 8160 SKX nodes.
	Top right: Data-parallel rendering of the Cosmic Web~\cite{iliev_simulating_2008} (29B transparent spheres),
	2FPS at $2048^2$ using 128 \theta \INTELXEONPHI Knight's Landing (KNL) nodes.
	Bottom: Data-parallel rendering of the 951GB DNS volume~\cite{lee_direct_2015} combined with a transparent
	isosurface (4.35B triangles), 5FPS at $4096\times1024$
	using 64 \stampede \INTELXEONPHI KNL nodes.}
}

\maketitle

\begin{abstract}
	Image- and data-parallel rendering across multiple
	nodes on high-performance computing systems is widely used in visualization
    to provide higher frame rates, support large data sets, and
    render data in situ.
	Specifically for in situ visualization, reducing bottlenecks incurred by the
	visualization and compositing is of key concern to reduce the overall
	simulation runtime.
    Moreover, prior algorithms have been designed to support either
    image- or data-parallel rendering
	and impose restrictions on the data distribution, requiring
different implementations for each configuration. In this paper,
we introduce the Distributed FrameBuffer, an asynchronous
image-processing framework for multi-node rendering.
We demonstrate that our approach achieves performance superior to
the state of the art for common use cases,
while providing the flexibility to support a wide range of parallel
    rendering algorithms and data distributions.
By building on this framework, we extend the open-source ray tracing library OSPRay with a data-distributed API, enabling
its use in data-distributed and in situ visualization applications.

\end{abstract}

\section{Introduction}
The need for high-performance distributed parallel rendering is growing,
spurred by trends in increasing data set sizes, the desire for
higher fidelity and interactivity, and the need for in situ visualization.
Meeting these demands poses new challenges to existing rendering methods,
requiring scalability across a spectrum of memory and compute capacities on
high-performance computing (HPC) resources.
Whereas the growth in data set sizes demands a large amount of aggregate
memory, the desire for more complex shading and interactivity
demands additional compute power. A large number
of application needs fall somewhere in between these extremes,
requiring a combination of additional
memory and compute. Finally, in situ visualization requires
the renderer to scale with the simulation, while incurring little
overhead.
Rendering techniques that scale well for either compute power or
aggregate memory capacity are well known, but applications falling
between these extremes have not been well addressed.



In large-scale rendering workloads on distributed-memory clusters,
the data is typically partitioned into subregions and distributed across
multiple nodes to utilize the aggregate memory available.
Each node is then responsible for rendering its assigned subregion of data.
The partial images rendered by each node
are then combined using a sort-last compositing
algorithm, e.g., Parallel Direct Send~\cite{hsu_segmented_1993},
Binary Swap~\cite{ma_parallel_1994}, Radix-k~\cite{peterka_configurable_2009},
or TOD-tree~\cite{grosset_tod-tree_2017}. The IceT library~\cite{moreland_image_2011}
provides implementations of a number of sort-last compositing algorithms
and is widely used in practice.
However, such data-parallel renderers impose restrictions on how the data
can be distributed, are susceptible to load imbalance,
and are limited to local illumination effects.

At the other end of the spectrum, the master-worker architecture has
been widely used to scale up
compute capacity and provide interactivity for high-fidelity
visualization of moderately sized data sets.
Master-worker renderers distribute work image-parallel by
assigning subregions of the image to be rendered by different nodes.
This architecture has been used effectively in a number of ray tracers,
e.g., Manta~\cite{bigler_design_2006}, OpenRT~\cite{wald_flexible_2002},
and OSPRay~\cite{wald_ospray_2017}.
While typically used for data which can be stored in memory on each node,
this architecture can be used for large data sets
by streaming data needed for the portion of the image
from disk~\cite{wald_interactive_2001-1} or over
the network~\cite{demarle_memory_2005,ize_real-time_2011};
however, these systems can suffer from cache thrashing
and are tied to specific data types or formats.



Applications falling somewhere in between the extrema of only
compute or memory scaling, or those seeking to go beyond common
use cases, can quickly run into issues with existing approaches.
For example, whereas a master-worker setup is well suited to
image-parallel ray tracing, if the renderer wants to perform additional
post-processing operations (e.g., tone-mapping, progressive refinement), or
handle multiple display destinations (e.g., display walls), the master
rank quickly becomes a bottleneck.
Similarly, whereas existing sort-last compositing algorithms are well
suited to statically partitioned data-parallel rendering, extending them
to support partially replicated or more dynamic data distributions
for better load balancing is challenging. Standard sort-last compositing
methods operate bulk-synchronously on the entire frame,
and are less suited to tile-based ray tracers in which small tiles are
rendered independently in parallel.

In this paper, we describe the algorithms and software architecture---the
``Distributed FrameBuffer''---that we developed to support distributed parallel
rendering, with the goal of addressing the above issues to
provide an efficient and highly adaptable framework suitable for a range of applications.
The Distributed FrameBuffer (DFB) is built on a tile-based work distribution
of the image processing tasks required to produce the final image
from a distributed renderer. These tasks are constructed per-tile at
runtime by the renderer and are not necessarily tied to the host application's
work or data distribution, providing the flexibility to implement a wide range
of rendering algorithms and distribute compute-intensive image processing tasks.
The DFB performs all
communication and computation in parallel with the renderer
using multiple threads 
to reduce compositing overhead.
Although the DFB is flexible enough to support renderers across
the spectrum of memory and compute scaling, it does not make a
performance trade-off to do so.
Our key contributions are:
\begin{itemize}
	\item A flexible and scalable parallel framework to execute
		compositing and image processing tasks for distributed rendering;
	\item A set of parallel rendering algorithms built on this approach,
		covering both standard use cases and more complex configurations;
	\item An extension of OSPRay to implement a data-distributed API,
		allowing end users to leverage the above algorithms in practice on
        a wide range of different data types.
\end{itemize}

\section{Previous Work}
\begin{figure*}
	\vspace{-0.5em}
	\centering
	\includegraphics[width=0.85\textwidth]{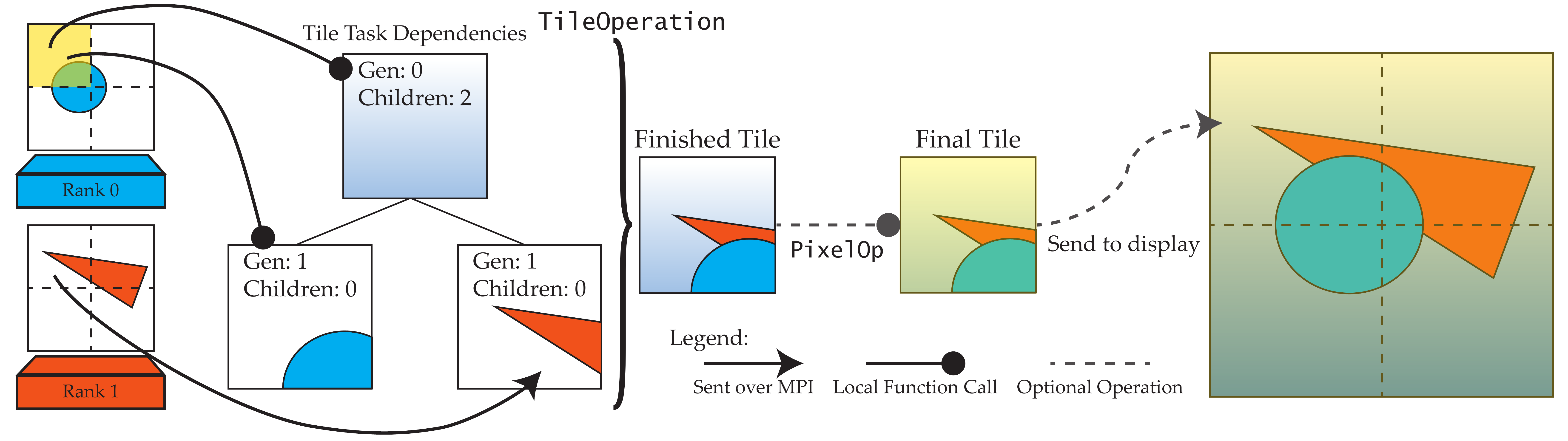}
	\vspace{-1em}
	\caption{\label{fig:dfb-tile-pipeline}%
	An example of the Distributed FrameBuffer's tile processing pipeline
	in a data-parallel renderer. Dependencies are specified
	on the fly per-tile and can be extended by child tiles.
	To compute the highlighted tile
	owned by rank 0, the owner sends
	a background color tile for generation 0, which specifies that two additional tiles
	will arrive in generation 1, potentially from different ranks.
	After receiving the full dependency tree, the tile operation
	produces the finished tile, which is tone-mapped by
	a pixel operation and sent to the display rank.}
	\vspace{-1.5em}
\end{figure*}


A large body of previous work has studied parallel rendering techniques for distributed-memory systems.
These works can generally be classified as one of three techniques, first discussed in the context of rasterization by Molnar et al.~\cite{molnar_sorting_1994}: sort-first, sort-middle, and sort-last.
Sort-middle is tied to rasterization, thus we focus our discussion on sort-first and sort-last strategies. 
Sort-first is an image-parallel technique, where the workload is distributed across multiple ranks by
subdividing the image.
Sort-last is a data-parallel technique, where the workload is distributed by
subdividing the 3D data, regardless of where it lies in the image.
Hybrid approaches have also been proposed, which combine sort-first and sort-last techniques.

\subsection{Data-Parallel Rendering}


In sort-last, or data-parallel, rendering the geometric primitives and volumetric data are partitioned in 3D space,
with each node assigned a subregion of the overall data to render.
In early implementations, this subdivision was at the level of a single primitive~\cite{roman79}.
Each node then renders its subregion of data to produce a partial image,
which must then be combined with other nodes' images to create the final image.
Combining these partial images typically requires depth compositing the
overlapping partial images to produce a correct final image.
It is this second step that becomes the bottleneck at large core counts
and high-resolutions, and therefore has been the focus of a large body of
work (e.g.,~\cite{hsu_segmented_1993,ma_parallel_1994,
favre_direct_2007,yu_massively_2008,kendall_accelerating_2010,moreland_image_2011,
grosset_dynamically_2016,grosset_tod-tree_2017}).

Most similar to our work in
the context of data-parallel rendering is Grosset et al.'s~\cite{grosset_dynamically_2016}
Dynamically Scheduled Region-Based compositing (DSRB)\@.
DSRB divides the image into strips and constructs a per-strip blending order,
referred to as a chain, based on which node's data projects to each strip.
Partial compositing for a chain can be done
after receiving strips from successive nodes in the chain,
overlapping compositing with rendering on other nodes.
However, DSRB is restricted in the amount of rendering it can overlap with compositing,
as each node renders its entire image before starting compositing;
is only applicable to data-parallel rendering;
and relies on a central scheduler to construct the chains.


The IceT library~\cite{moreland_image_2011} encompasses several different compositing strategies for sort-last rendering
and has been widely deployed across popular parallel scientific visualization tools.
Thus, we use IceT as a primary point of comparison when evaluating our method's performance.
Although IceT was initially designed for rasterization, Brownlee et al.~\cite{brownlee_study_2012} used IceT's depth compositing
with a ray tracer inside of multiple visualization tools, though were hindered by the data distribution
chosen by the tools. Wu et al.~\cite{wu_visit-ospray_2018} employed a similar approach to integrate OSPRay
into VisIt, using OSPRay to render locally on each rank and IceT to composite the image, and encountered
similar difficulties.

\subsection{Image-Parallel Rendering}
Image-parallel renderers assign subregions of the image to different ranks
for rendering. To render large datasets, this approach is typically coupled with some form
of data streaming or movement into a local cache, and is designed to exploit frame-to-frame coherence.
The data movement work is amortized over multiple frames as the data rendered for a region of
the image in one frame will likely be similar to that rendered in the next frame.
Early rasterization-based techniques used a sort-middle algorithm, where the image was partitioned between nodes, and geometry
sent to the node rendering the portion of the image it projected to~\cite{ellsworth90}.

Image-parallel rendering lends itself well to ray tracing, as ray tracers already use acceleration structures
for ray traversal which can be readily adapted to streaming and caching portions of the scene as they are traversed. 
Wald et al.~\cite{wald_interactive_2001-1} used a commodity cluster for interactive ray tracing of large
models, where a top-level $k$-d tree is replicated across the nodes and lower sub-trees
fetched on demand from disk.
DeMarle et al.~\cite{demarle_memory_2005}
used an octree acceleration structure for rendering large volume data, where missing voxels
would be fetched from other nodes using a distributed shared memory system.
Ize et al.~\cite{ize_real-time_2011} extended this approach to
geometric data using a distributed BVH\@. When rendering fully replicated
data, their approach avoids data movement and compositing, and can achieve
100FPS for primary visibility ray casting on 60 nodes.  Biedert et 
al.~\cite{biedert18} proposed an image-parallel remote streaming framework
able to achieve over 80FPS from a distributed cluster to a remote client,
using hardware acceleration and adaptive tile-based streaming.



\subsection{Hybrid-Parallel Rendering}
While image- and data-parallel rendering methods distribute work solely by partitioning
the image or data, hybrid-parallel renderers combine both strategies, aiming to pick
the best for the task at hand. Reinhard et al.~\cite{reinhard_hybrid_1999}
first proposed a hybrid scheduling algorithm for ray tracing distributed data, where
the rays would be sent or the required data fetched depending on the coherence
of the rays.

Samanta et al.~\cite{samanta_hybrid_2000} proposed to combine
sort-first and sort-last rendering in the context of a rasterizer, by partitioning
both the image and data among the nodes. Each node then renders its local data and sends
rendered pixels to other nodes that own the tiles its data touches. The tiles are then
composited on each node and sent to the display node. This approach
bears some resemblance to the Distributed FrameBuffer, although lacks its
extensibility and support for ray tracing specific rendering effects.


Navr\'atil et al.~\cite{navratil_dynamic_2012} proposed a scheduler that combines
static image and data decompositions for ray tracing, roughly similar to
sort-first and sort-last, respectively. However, a key difference of their approach when
compared to a sort-last rasterizer is that rays will be sent between nodes, similar to
Reinhard et al.~\cite{reinhard_hybrid_1999}, to compute reflections and shadows.
The static image decomposition scheduler works similar to the image-parallel
algorithms discussed previously.
Abram et al.~\cite{abram_galaxy_2018} extended the domain
decomposition model to an asynchronous, frameless renderer using a subtractive lighting
model for progressive refinement.
Park et al.~\cite{park_spray_2018} extended both the image and domain decompositions,
by introducing ray speculation to improve system utilization and overall performance.
By moving both rays or data as needed, these approaches are able
to compute global illumination effects on the distributed data, providing
high-quality images at additional cost.

Biedert et al.~\cite{biedert_task-based_2017} employed a task-based model
of distributed rendering which is able to combine sort-first and sort-last
rendering, by leveraging an existing runtime system to balance between these
strategies. Although their work uses OSPRay for rendering, it is restricted to a single
thread per-process and is non-interactive.

\subsection{OSPRay, Embree and ISPC}
Although the Distributed FrameBuffer is applicable to any
tile-based rendering algorithm, we evaluate it within
the context of the OSPRay ray tracing framework~\cite{wald_ospray_2017}.
OSPRay provides a range of built in volume and geometric primitives
used in scientific visualization, advanced shading
effects, and achieves interactive rendering on typical workstations and laptops.
To achieve interactive ray tracing performance on CPUs,
OSPRay builds on top of Embree~\cite{wald_embree_2014},
the Intel SPMD Program
Compiler (ISPC)~\cite{pharr_ispc_2012},
and Intel's Threading Building Blocks (TBB)\@.

Embree is a high-performance kernel
framework for CPU ray tracing, and provides a set of low-level kernels
for building and traversing ray tracing data structures which
are highly optimized for modern CPU architectures.
ISPC is a single program multiple data (SPMD) compiler,
which vectorizes a scalar program by mapping different instances of
the program to the CPU's vector lanes, thereby executing them in parallel.
TBB provides a set of parallel programming primitives for writing
high-performance multi-threaded code, similar to OpenMP\@.


\section{The Distributed FrameBuffer}
\label{sec:the-dfb}
At its core, the Distributed FrameBuffer (DFB) is not a
specific compositing algorithm per se, but a general framework
for distributed rendering applications.
A renderer using the DFB specifies a set of
tasks to be executed on the rendered image and
per-tile dependency trees for the tasks.
The tasks are parallelized over the image by subdividing it into tiles,
where each tile is owned by a unique rank---the tile owner---responsible
for executing tasks for that tile.
If task dependencies are produced on ranks other
than the tile owner the DFB will route them over the network to the owner.
The tile dependency trees are specified per-tile and per-frame,
allowing for view- and frame-dependent behavior.

The tile processing pipeline involves three stages (\Cref{fig:dfb-tile-pipeline}).
First, the dependency tree is constructed by the tile operation
as task dependencies are received from other ranks.
Once the entire tree has been received the finished tile is computed
by the tile operation and passed on to any pixel operations.
The final output tile is then converted to the display image format
and sent to the display rank, if needed.
The processing pipeline and messaging system run
asynchronously on multiple threads,
allowing users to overlap additional
computation with that performed by the DFB\@.
Although the focus of this paper is on using the DFB for rendering,
the task input tiles are not required to be produced
by a renderer.


\subsection{Tile Processing Pipeline}
The DFB begins and ends processing synchronously, allowing applications
processing multiple frames, i.e., a renderer, to ensure that tiles
for different frames are processed in the right order.
Before beginning a frame, the renderer specifies the tile operation
to process the tiles it will produce.
Each rank then renders some set of tiles
based on the work distribution chosen by the renderer.
As tiles are finished, they are handed to the DFB for processing
by calling \texttt{setTile}. During the frame, the DFB will compute
tile operations for the tiles owned by each rank in the background
and send other tiles over the network to their owner.
The frame is completed on each rank when the tiles it owns
are finalized, and rendering is finished when all processes
have completed the frame. As each tile is processed independently in parallel
it is possible for some tiles to be finalized while
others have yet to receive their first inputs.


To track the distributed tile ownership, the DFB instance on each rank stores a
tile descriptor (\Cref{lst:tile-structs}) for each tile in the image.
When \texttt{setTile} is called
the DFB looks up the descriptor for the tile and sends it
to the owner using an asynchronous messaging layer (\Cref{sec:dfb-messaging-layer}).
If the owner is the calling rank itself, the tile is instead scheduled for processing
locally.

\begin{listing}[t]
\begin{minted}[mathescape,fontsize=\scriptsize]{c++}
struct Tile {
    int generation;
    int children;
    region2i screenRegion; 
    int accumulationID; // Sample pass for progressive refinement
    float color[4*TILE_SIZE*TILE_SIZE];
    float depth[TILE_SIZE*TILE_SIZE];
    float normal[3*TILE_SIZE*TILE_SIZE]; // Optional
    float albedo[3*TILE_SIZE*TILE_SIZE]; // Optional
};
struct TileDescriptor {
    virtual bool mine() { return false; }
    vec2i coords;
    size_t tileID, ownerRank;
};
struct TileOperation : TileDescriptor {
    bool mine() { return true; }
    virtual void newFrame() = 0;
    virtual void process(const Tile &tile) = 0;

    DistributedFrameBuffer *dfb;
    vec4f finalPixels[TILE_SIZE*TILE_SIZE];
    Tile finished, accumulation, variance;
};
\end{minted}
	\vspace{-1.5em}
	\caption{\label{lst:tile-structs}%
	The base structures for tiles and tile operations.\vspace{-2em}}
\end{listing}

For each tile owned by the rank, the DFB stores a
concrete tile operation instance in the array of descriptors.
The base structure for tile operations (\Cref{lst:tile-structs})
stores a pointer to the local DFB instance and
a Tile buffer to write the finished tile data to,
along with optional accumulation and variance buffer tiles.
The \texttt{finalPixels} buffer is used as scratch space to
write the final tile to, before sending it to the display rank.

To implement the corresponding tile operation for a rendering algorithm
(e.g., sort-last compositing) users extend the \texttt{TileOperation},
and specify their struct to be used by the DFB\@.
Each time a tile is received by the DFB instance on the tile owner, the \texttt{process}
function is called on the tile operation to execute the task.
The \texttt{newFrame} function is called when a new frame begins,
to reset any per-frame state.

When all a tile's dependencies have been received the tile operation
combines the inputs to produce a finished tile, which is then passed to the DFB\@.
The local DFB instance runs any
additional pixel operations on the finished tile and converts the final pixels
to the display color format, outputting them to the
\texttt{finalPixels} buffer. This buffer is then compressed and sent to
the display rank. In addition to the \texttt{RGBA8} and
\texttt{RGBAF32} display formats, the DFB also offers a \texttt{NONE}
format, which is unique in that it indicates that the display
rank should not receive or store the final pixels at all.
We will discuss a useful application of the \texttt{NONE} format
in \Cref{sec:dfb-apps-display-wall}.


\subsubsection{Per-Tile Task Dependency Trees}
\label{sec:tile-task-deps}
The \texttt{Tile} structure passed to \texttt{setTile} and routed
over the network is shown in \Cref{lst:tile-structs}.
To construct the dependency tree, each rendered tile specifies itself
as a member of some generation (a level in the tree), and as having
some number of children in the following generation.
The total number of tiles to expect in the next generation is the
sum of all children specified in the previous one.
Different ranks can contribute tiles with
varying numbers of children for each generation, and can send
child tiles for parents rendered by other ranks.
There is no requirement that tiles are sent in order by generation,
nor is a tile operation guaranteed to receive tiles in a
fixed order. Tile operations with dependencies beyond a trivial
single tile can be implemented by buffering received tiles in \texttt{process}
to collect the complete dependency tree.

The interpretation and processing order of the dependency tree
is left entirely to the tile operation.
For example, the dependency tree could be used
to represent a compositing tree, input to some filtering,
or simply a set of pixels to average together.
The creation of the dependency trees by the renderer
and their processing by the tile operation are tightly coupled,
and thus the two are seen together as a single
distributed rendering algorithm.
The flexibility of the tile operation and dependency trees
allows the DFB to be used in a wide range of rendering
applications (\Cref{sec:dfb-apps}).

\subsubsection{Pixel Operations}
\label{sec:dfb-pixel-ops}
Additional post-processing, such as tone-mapping, can be performed
by implementing a pixel operation (\texttt{PixelOp}).
The pixel operation takes the single finished tile from the
tile operation as input, and thus is not tied to the tile operation
or renderer. The DFB runs the pixel operation on the tile owner
after the tile operation is completed to distribute the work.
In addition to image post-processing, pixel operations can be used,
for example, to re-route tiles to a display
wall (\Cref{sec:dfb-apps-display-wall}).

\subsection{Asynchronous Messaging Layer}
\label{sec:dfb-messaging-layer}
To overlap communication between nodes with computation,
we use an asynchronous point-to-point
messaging layer built on top of MPI (Message Passing Interface).
Objects that will send and receive messages register themselves
with the messaging layer and specify a unique global identifier.
Each registered object is seen as a global ``distributed object'', with
an instance of the object on each rank which can be looked up
by its global identifier.
A message can be sent to the instance of an object on some
rank by sending a message to the rank with
the receiver set as the object's identifier.

The messaging layer runs on two threads:
a thread which manages sending and receiving messages with MPI,
and an inbox thread which takes received messages and passes them
to the receiving object.
Messages are sent by pushing them on to an outbox,
which is consumed by the MPI thread.
To avoid deadlock between ranks,
we use non-blocking MPI calls
to send, receive, probe, and test for message completion.
Received messages
are pushed on to an inbox, which is consumed by the
inbox thread. To hand a received message to the receiving object,
the inbox thread looks up the receiver by its global ID in a hash table.
Messages are compressed using
Google's Snappy library~\cite{google_snappy_nodate} before enqueuing
them to the outbox and decompressed on the inbox thread before being
passed to the receiver.

In our initial implementation we also used the messaging layer to gather
the final tiles to the display rank. However, unless the
rendering workload is highly imbalanced, this approach generates a
large burst of messages to the display, with
little remaining rendering work to overlap with.
This burst of messages also appeared to trigger an
MPI performance issue on some implementations.
As an optimization,
the final tiles are instead written to a buffer, which is compressed
and gathered to the display with a single \texttt{MPI\_Gatherv}
at the end of the frame.

\section{Rendering with the Distributed FrameBuffer}
\label{sec:dfb-applications}
\label{sec:dfb-apps}
%

A distributed rendering algorithm using the DFB consists of
a renderer, responsible for rendering tiles
of the image, coupled with a tile operation, which will combine
the results of each ranks' renderer.
In the following sections we discuss a few distributed rendering
algorithms built on the DFB,
covering standard image-parallel
(\Cref{sec:dfb-apps-img-parallel}) and
data-parallel (\Cref{sec:dfb-apps-data-parallel}) rendering,
along with extensions to these methods enabled by the
DFB, specifically, dynamic load balancing (\Cref{sec:dfb-apps-load-balancing})
and mixed-parallel rendering (\Cref{sec:dfb-apps-hybrid-parallel}).
Finally, we discuss how pixel operations can be used to implement a
high-performance display wall system (\Cref{sec:dfb-apps-display-wall}).

\subsection{Image-Parallel Rendering}
\label{sec:dfb-apps-data-replicated}
\label{sec:dfb-apps-img-parallel}
An image-parallel renderer distributes
the tile rendering work in some manner between the ranks
such that each tile is rendered once.
This distribution can be a simple linear assignment,
round-robin, or based on some runtime load balancing.
The corresponding tile operation
expects a single rendered tile as input.
The DFB allows for a straightforward and elegant
implementation of this renderer (\Cref{lst:image-parallel-renderer}).


\begin{listing}[t]
\begin{minted}[mathescape,fontsize=\scriptsize]{c++}
struct ImageParallel : TileOperation {
    void process(const Tile &tile) {
        // Omitted: copy data from the tile
        dfb->tileIsCompleted(this);
    }
};
void renderFrame(DFB *dfb) {
    dfb->begin();
    parallel_for (Tile &t : assignedTiles()) {
        renderTile(t);
        dfb->setTile(t);
    }
    dfb->end();
}
\end{minted}
	\vspace{-0.5em}
	\caption{\label{lst:image-parallel-renderer}%
	The tile operation and rendering loop for an
	image-parallel renderer using the DFB\@.\vspace{-1.5em}}
\end{listing}

\subsubsection{Tile Ownership vs.\ Work Distribution}
\label{sec:dfb-apps-load-balancing}
The work distribution chosen by the renderer
is not tied to the DFB tile ownership, allowing the renderer
to distribute work as desired. Though it is preferable
that the tile owners render the tiles they own
to reduce network traffic, this is not a requirement.


This flexibility in work distribution can be used, for example,
to implement dynamic load balancing. We extend the \texttt{ImageParallel}
tile operation to support receiving a varying number of tiles,
and the renderer to assign each tile to multiple ranks.
Each redundantly assigned tile uses a different random seed to generate camera rays,
thereby computing a distinct set of samples.
The rendered tiles are then averaged together by the tile operation,
producing a finished tile equivalent to a higher sampling rate.
This approach is especially
useful for path tracing, as a high number of samples
are required to produce a noise-free image. Tiles with higher variance
can be assigned to additional ranks, adjusting the sampling rate dynamically.


\subsection{Data-Parallel Rendering}
\label{sec:dfb-apps-data-parallel}
A standard sort-last data-parallel renderer decomposes the scene into a set of bricks,
and assigns one brick per-rank for rendering. Each rank renders its local data to
produce a partial image, which are combined using
a sort-last compositing algorithm to produce an image of the entire dataset.
To implement a data-parallel renderer using the DFB, we
express sort-last compositing as a tile operation, and take advantage of the DFB's
asynchronous tile routing and processing to execute the compositing
in parallel with local rendering.
The benefits of this approach are two-fold:
the per-tile task dependencies allow to minimize compositing and communication
work per-tile, and
overlapping compositing and rendering reduces the additional time
spent compositing after rendering is finished.

\begin{figure}[t]
	\includegraphics[width=0.95\columnwidth]{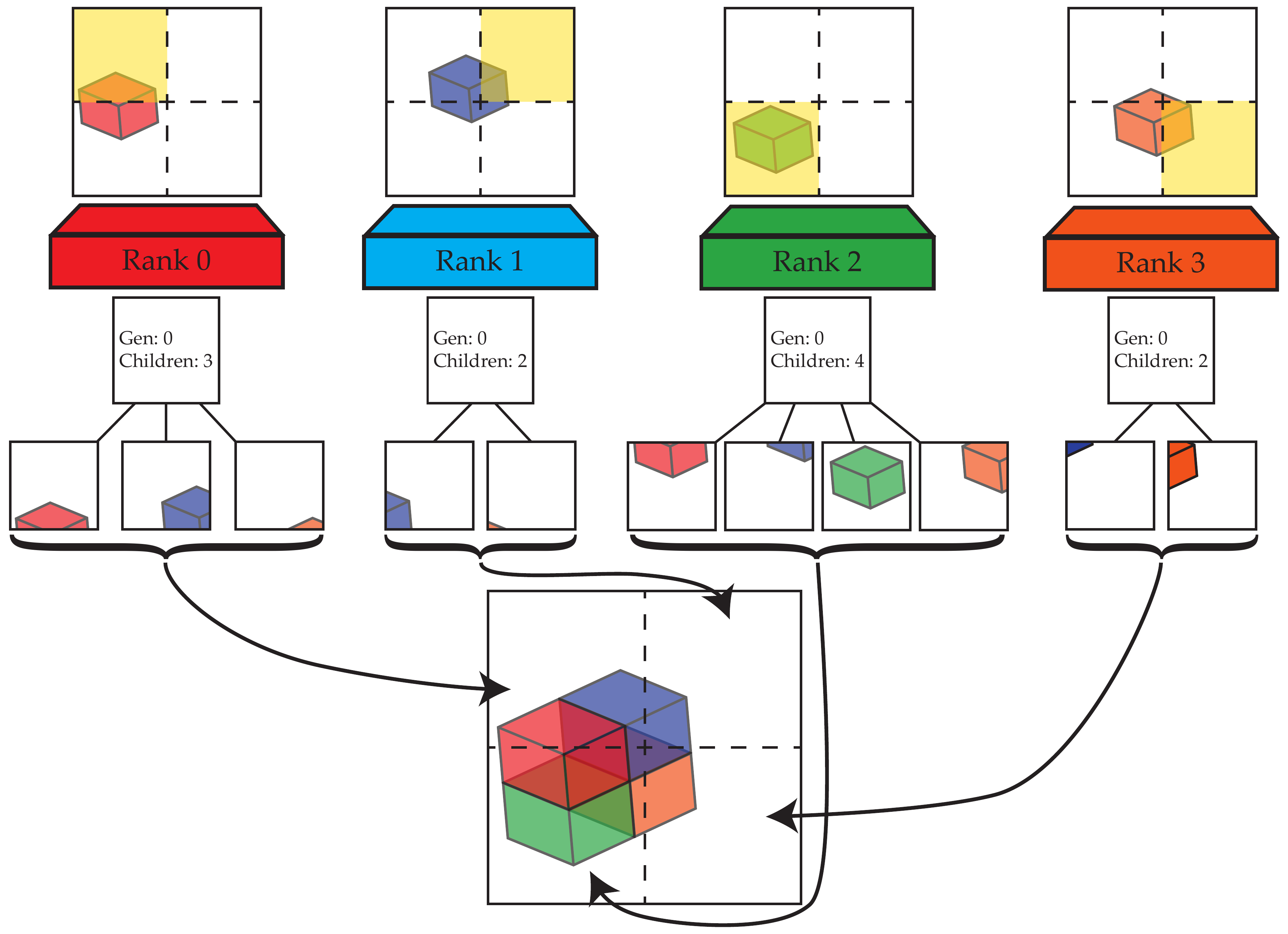}
	\vspace{-1em}
	\caption{\label{fig:dfb-simple-data-parallel}%
	Tile ownership and dependency trees for a data-parallel
	renderer using the DFB. Each rank owns its highlighted tile, and
	receives input tiles from ranks whose data projects
	to the tile. Compositing runs
	in parallel to local rendering, reducing overhead.}
	\vspace{-0.5em}
\end{figure}


To compute a per-tile compositing dependency tree, each rank collects the bounds
of the other ranks' data and projects them to the image
(\Cref{fig:dfb-simple-data-parallel}). Only those ranks whose data
projects to some tile will render inputs for it.
Each rank is responsible for specifying the dependency information
for the tiles it owns (highlighted in yellow, \Cref{fig:dfb-simple-data-parallel}).
The tile owner
will compute an additional ``background'' tile and
set it as the sole member of generation 0.
The background tile is filled with the background
color or texture, and sets the number of ranks whose data
project to the tile as the number of children.


\begin{listing}[t]
\begin{minted}[mathescape,linenos,fontsize=\scriptsize]{c++}
void renderFrame(Brick local, box3f allBounds[], DFB *dfb) {
    dfb->begin();
    /* We touch the tiles we own and those touched by the
       screen-space projection of our brick */
    Tile tiles[] = {dfb->ownedTiles(), dfb->touchedTiles(local)};
    parallel_for (Tile &t : tiles) {
        bool intersected[] = intersectBounds(allBounds, t);
        if (dfb->tileOwner(t)) {
            fillBackground(t);
            t.generation = 0;
            t.children = numIntersected(intersected).
            dfb->setTile(t);
        }
        if (intersected[local]) {
            renderBrickForTile(t, local);
            t.generation = 1;
            t.children = 0;
            dfb->setTile(t);
        }
    }
    dfb->end();
}
\end{minted}
	\vspace{-0.75em}
	\caption{\label{lst:data-parallel}%
	The rendering loop for a standard data-parallel renderer.\vspace{-1.5em}}
\end{listing}

The renderer (\Cref{lst:data-parallel}) begins by determining the
set of candidate tiles that it must either send a background tile for
or render data to.
The candidate tiles that the rank's local data may project to
are found using a conservative screen-space AABB test, which is subsequently refined.
For each candidate tile, the renderer computes an exact list of the
ranks whose data touches the tile by ray tracing the bounding boxes.
The number of intersected boxes is the number of generation 1
tiles to expect as input to the tree.
If the rank's local data was intersected, it renders its
data and sends a generation 1 tile.
To allow for ghost zones and voxels, camera rays
are clipped to the local bounds of the rank's data.
As with the outer candidate tile loop,
the inner rendering loop is parallelized over the pixels in a tile.

After receiving the entire dependency tree, the \texttt{AlphaBlend} tile operation
(\Cref{lst:alphablend-tileop}) sorts the pixels by depth and blends them
together to composite the tile.
The tile fragment sorting is done per-pixel, in contrast to the per-rank sort
used in standard approaches. Sorting per-pixel allows for
rendering effects like depth of field, side-by-side stereo, and dome projections.
As the tile processing is done in parallel, we do not find
the sorting to be a bottleneck.
In the case that a rank-order sort would produce a correct image,
the dependency tree can be constructed as a list instead
of a single-level tree with tiles ordered back-to-front
by generation.
Finally, although we have discussed the
data-parallel renderer with a single brick of data per-rank, it trivially supports
multiple bricks per-rank, allowing for finer-grained
work distributions.

\begin{listing}[t]
\begin{minted}[mathescape,fontsize=\scriptsize]{c++}
struct AlphaBlend : TileOperation {
    Tile bufferedTiles[];
    int currentGen, missing, nextChildren;
    void newFrame() {
        currentGen = 0;
        missing = 1; // Expect a generation 0 tile to start
        nextChildren = 0;
    }
    void process(const Tile &tile) {
        bufferedTiles.append(tile);
        if (tile.generation == currentGen) {
            --missing;
            nextChildren += tile.children;
            checkTreeComplete();
        }
        if (!missing) {
            sortAndBlend(bufferedTiles);
            dfb->tileIsCompleted(this);
            bufferedTiles = {}
        }
    }
    // Check receipt of all children from all generations,
    // advancing currentGen as we complete generations.
    void checkTreeComplete() { /* omitted for brevity */ }
}
\end{minted}
	\vspace{-0.5em}
	\caption{\label{lst:alphablend-tileop}%
	The sort-last compositing tile operation used by the data- and mixed-parallel
	renderers. It first collects the dependency tree,
	then sorts and blends the pixels to produce the
	composited tile.\vspace{-1.5em}}


\end{listing}

\subsection{Rendering Hybrid Data Distributions}
\label{sec:dfb-apps-hybrid-parallel}
\label{sec:dfb-apps-mixed-parallel}

A data-parallel renderer that statically assigns
each brick of data to a single rank is
susceptible to load imbalance, coming from factors
such as the data distribution, transfer function,
or camera position.
To better distribute the workload, we can 
assign the same brick of data to multiple ranks,
with each rank potentially assigned multiple bricks.
Each rank is responsible for rendering
a subset of the tiles the bricks it has projects to, thereby dividing
the rendering workload for each brick among the ranks.
Although this increases the memory requirements of the
renderer, additional memory is often available
given the number of compute nodes used to achieve an interactive frame rate.

Rendering such a configuration with a standard
compositing approach is either difficult or not possible, as the compositing tree
and blending order is set for the entire framebuffer by sorting the
ranks~\cite{moreland_image_2011}.
However, the DFB's per-tile dependency trees allow renderers to change which ranks contribute
tiles for each image tile.
This enables a direct extension of the data-parallel
renderer discussed previously into a mixed-parallel renderer,
which balances image and data parallelism to achieve better load balance.

To develop the mixed-parallel extension, we introduce the concept of
a ``tile-brick owner''. Given a dataset partitioned into a set of bricks
and distributed among the ranks with some level of replication, the
renderer must select a unique rank among those
sharing a brick to render it for each image tile.
The rank chosen to
render the brick for the tile is referred to as the ``tile-brick owner''.
Thus we can take our data-parallel renderer and
modify it so that a rank will render a brick for a tile if the
brick projects to the tile and the rank is the tile-brick
owner (\Cref{lst:hybrid-parallel}).
The task dependency tree and tile operation are the same 
as the data-parallel renderer; the only difference is
which rank renders the generation 1 tile for a given brick and image tile.

\begin{listing}[t]
\begin{minted}[mathescape,fontsize=\scriptsize]{c++}
void renderFrame(Brick local[], box3f allBounds[], DFB *dfb) {
    dfb->begin();
    Tile tiles[] = {dfb->ownedTiles(), dfb->touchedTiles(local)};
    parallel_for (Tile &t : tiles) {
        bool intersected[] = intersectBounds(allBounds, t);
        if (dfb->tileOwner(t)) {
            // Listing 3, lines 9-12
        }
        parallel_for (Brick &b : local) {
            if (tileBrickOwner(b, t) && intersected[b]) {
                // Listing 3, lines 15-18
            }
        }
    }
    dfb->end();
}
\end{minted}
	\vspace{-0.5em}
	\caption{\label{lst:hybrid-parallel}%
	The rendering loop of the mixed-parallel renderer.
	The DFB allows for an elegant extension of the data-parallel
	renderer to support partially replicated data for better load-balancing.\vspace{-1.5em}}
\end{listing}

Our current renderer
uses a round-robin assignment to select tile-brick ownership,
however this is not a requirement of the DFB\@.
A more advanced renderer could assign
tile-brick ownership based on some load-balancing strategy (e.g.,~\cite{frey_load_2011}),
or adapt the brick assignment based on load imbalance measured
in the previous frame.
The strategies discussed for image-parallel load balancing
and work subdivision in \Cref{sec:dfb-apps-load-balancing}
are also applicable to the mixed-parallel renderer. For example,
two ranks sharing a brick could each compute half of the camera rays
per-pixel, and average them together in the tile operation
to produce a higher quality image.

The mixed-parallel renderer supports the entire
spectrum of image- and data-parallel rendering: given a single
brick per-rank it is equivalent to the data-parallel
renderer; given the same data on all ranks
it is equivalent to the image-parallel renderer;
given a partially replicated set of data, or a mix of fully replicated
and distributed data, it falls in between.

\subsection{Display Walls}
\label{sec:dfb-apps-display-wall}

\begin{figure}
	\centering
	\includegraphics[width=0.85\columnwidth]{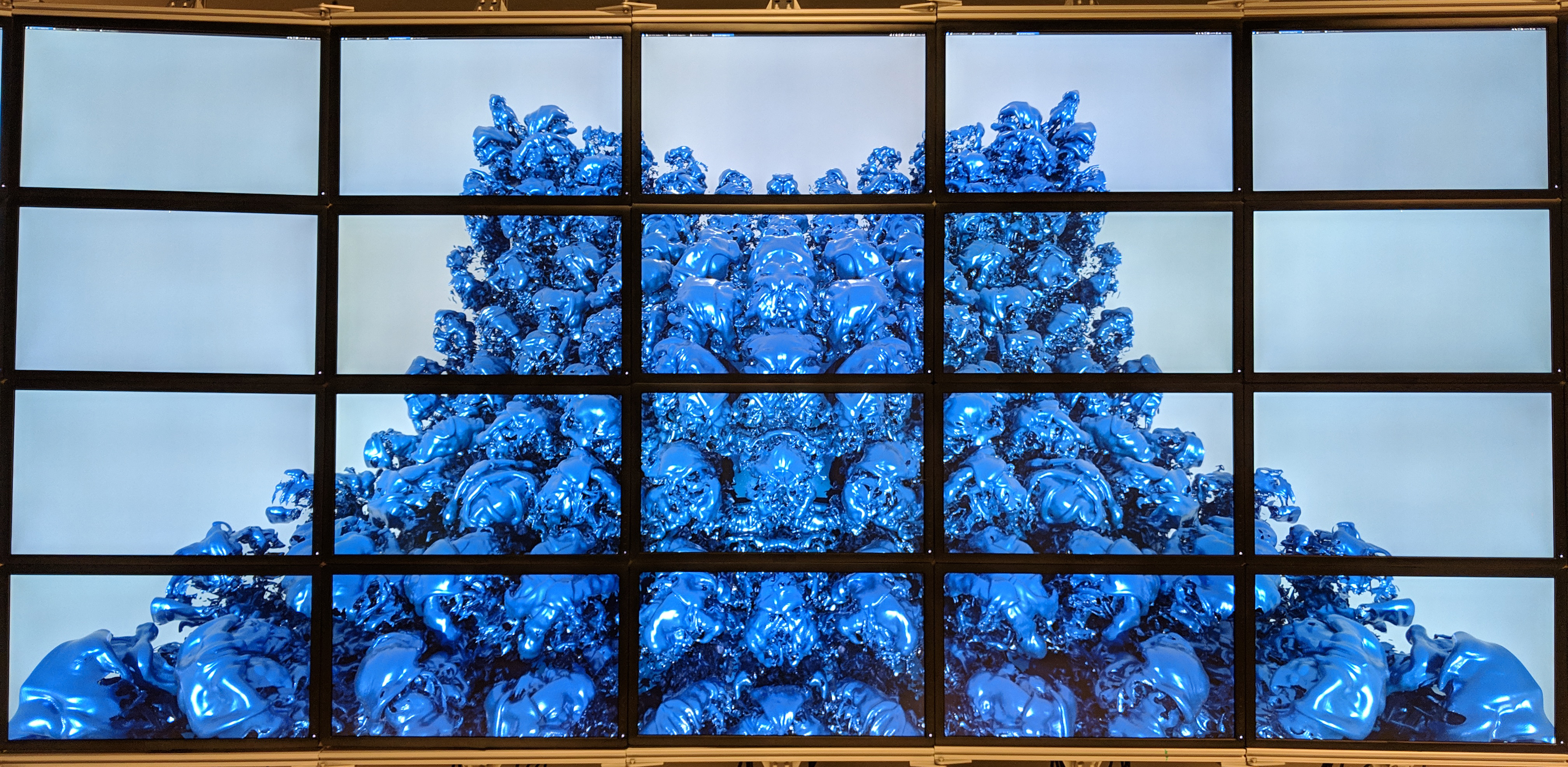}
	\vspace{-0.75em}
	\caption{\label{fig:display-wall}%
	A prototype display wall system using DFB pixel operations to
	send tiles in parallel from an image-parallel path tracer.}
	\vspace{-1.5em}
\end{figure}

The DFB can also be used to implement a high-performance display wall rendering
system by using a pixel operation to send tiles directly to the displays (\Cref{fig:display-wall}).
Tiles will be sent in parallel as they are finished on the tile owner directly to
the displays, achieving good utilization of a fully interconnected network. 
Moreover, when rendering with the \texttt{NONE} image format, the image
will not be gathered to the master rank, avoiding a large amount of
network communication and a common bottleneck. As pixel operations
are not tied to the rendering algorithm or tile operation, this method
can be used to drive a display wall with any of the presented
renderers.


\subsection{Implementation}
\label{sec:dfb-apps-implementation}
We implement the Distributed FrameBuffer and the presented rendering
algorithms in OSPRay's MPI module, using Intel TBB for
multi-threading and ISPC~\cite{pharr_ispc_2012} for vectorization.
The underlying
implementation of the \texttt{MPIDevice} provided by
OSPRay~\cite{wald_ospray_2017} for image-parallel rendering
has been significantly improved by this work, although
it is exposed to users in the same manner as before. Users can continue to run
existing OSPRay applications with \texttt{mpirun} and pass
the \texttt{-{}-osp:mpi} argument to the application, and OSPRay
will replicate the scene data across a cluster and render it
image-parallel using the rendering algorithms described
in \Cref{sec:dfb-apps-data-replicated,sec:dfb-apps-load-balancing}.

\begin{figure*}
	\centering
	\begin{subfigure}{0.24\textwidth}
		\centering
		\includegraphics[width=\textwidth]{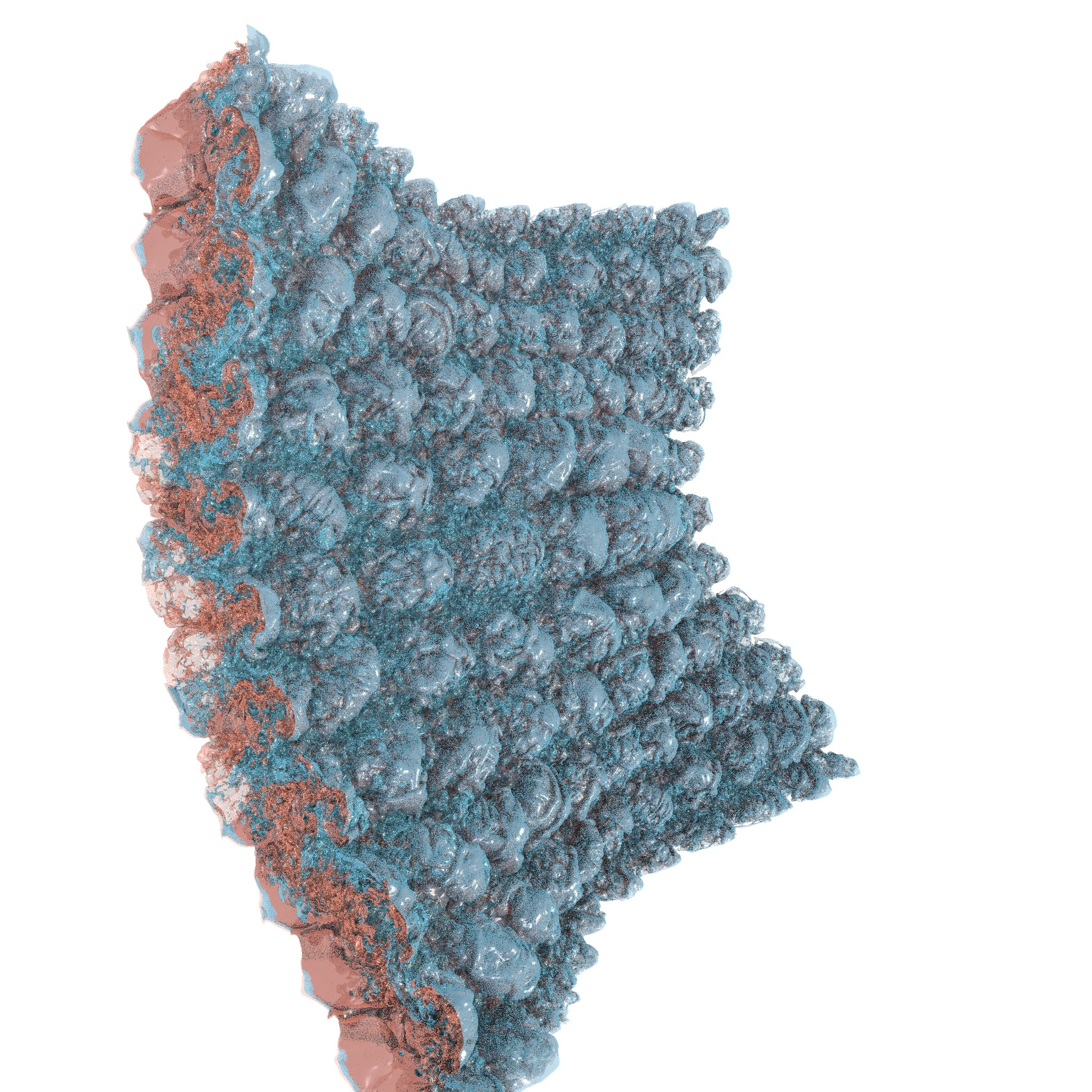}
		\caption{\label{fig:rm-iso-benchmark}R-M transparent isosurfaces.} 
	\end{subfigure}
	\begin{subfigure}{0.24\textwidth}
		\centering
		\includegraphics[width=\textwidth]{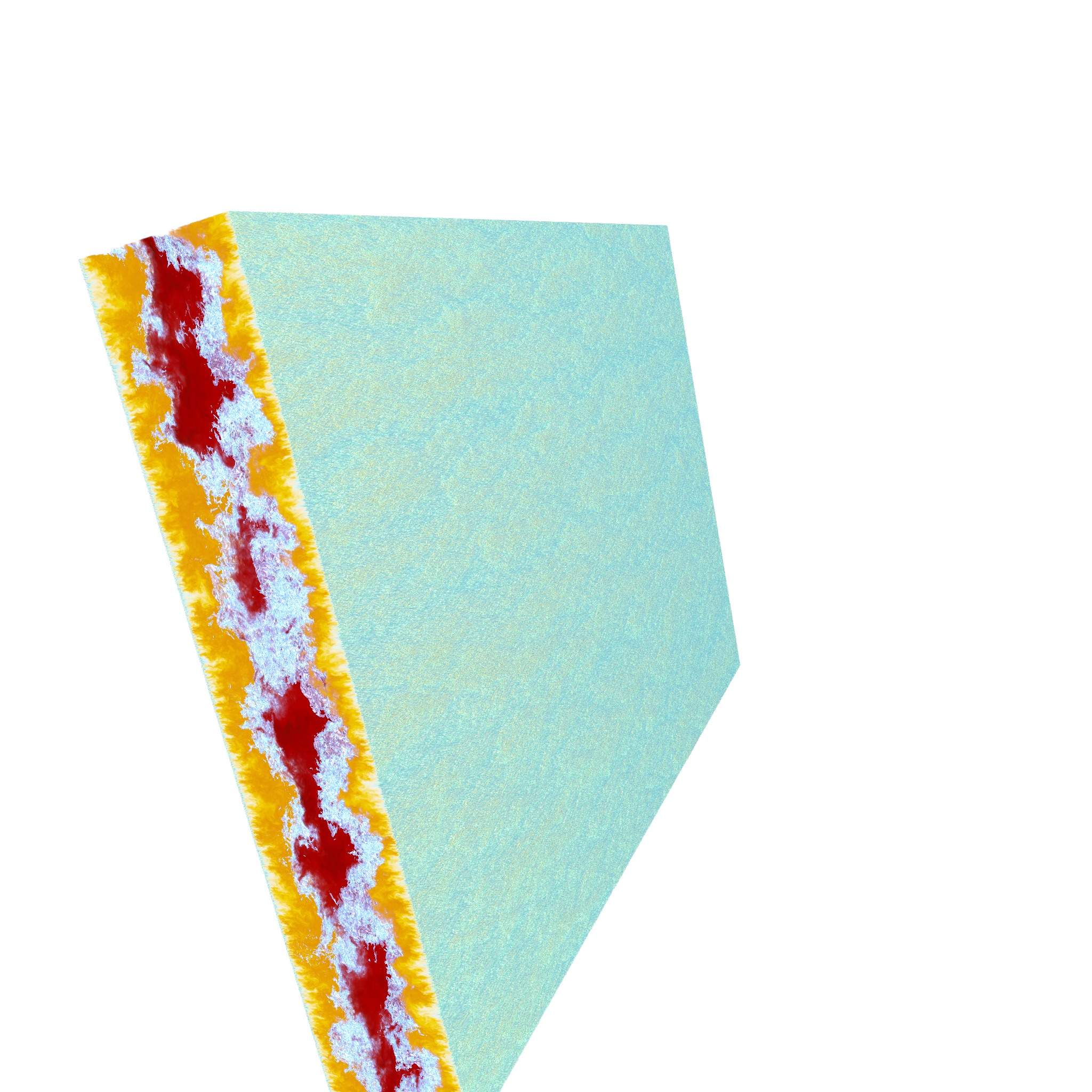}
		\caption{\label{fig:dns-iso-vol-benchmark}DNS with transparent isosurfaces.}
	\end{subfigure}
	\begin{subfigure}{0.24\textwidth}
		\centering
		\includegraphics[width=\textwidth]{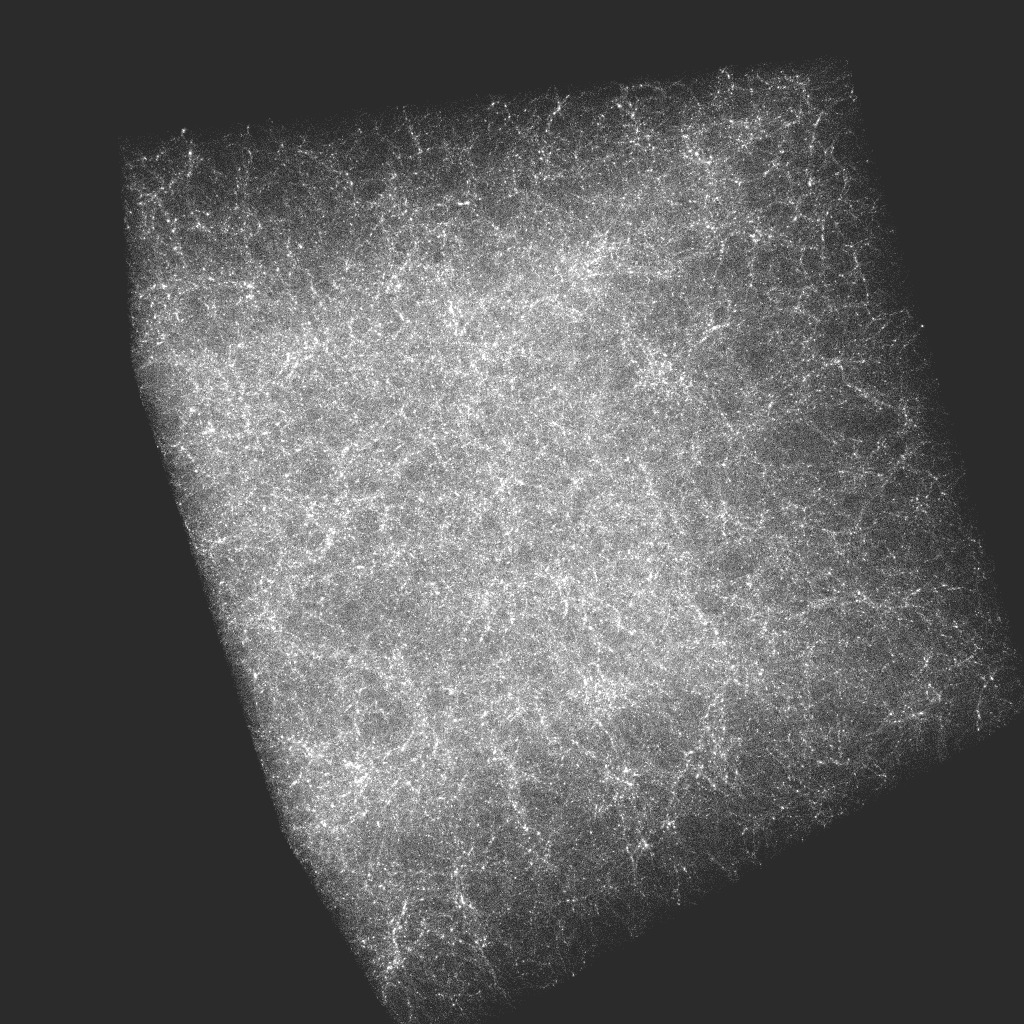}
		\caption{\label{fig:cosmic-web-benchmark}$5^3$ Cosmic Web subset.} 
	\end{subfigure}
	\begin{subfigure}{0.24\textwidth}
		\centering
		\includegraphics[width=\textwidth]{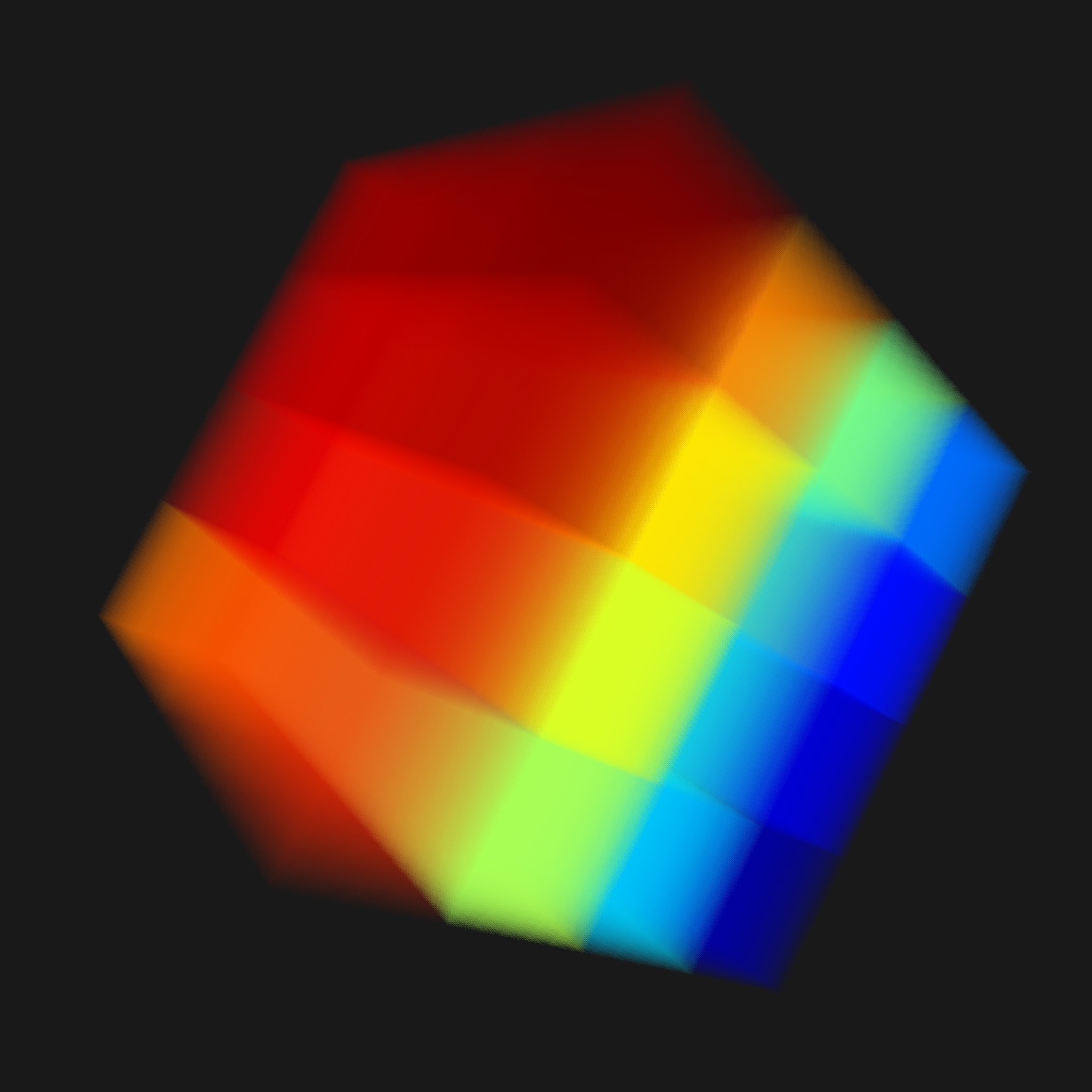}
		\caption{\label{fig:compositing-benchmark}%
		Synthetic benchmark volume.}
	\end{subfigure}
	\vspace{-1em}
	\caption{\label{fig:benchmark-datasets}%
	The data sets used in our benchmarks.
	(a) Two transparent isosurfaces on the Richtmyer-Meshkov~\cite{cohen_rm_2002}, 516M triangles total.
	(b) A combined visualization of the 451GB single-precision DNS~\cite{lee_direct_2015}
	with two transparent isosurfaces, 5.43B triangles total.
	(c) A $5^3$ subset of the $8^3$ Cosmic Web~\cite{iliev_simulating_2008},
	7.08B particles rendered as transparent spheres.
	(d) The generated volume data set used in the compositing benchmarks, shown for 64 nodes.
	Each node has a single $64^3$ brick of data.}
	\vspace{-1.5em}
\end{figure*}

\section{A Data-Distributed API for OSPRay}
\label{sec:ospray-dp-api}


The OSPRay API was originally designed for
a single application process passing its data to OSPRay. Although
OSPRay may offload the data in some way to other ranks, this is done without the
application's awareness.
This API works well for applications that do not
need to specify the data distribution; however, it is not applicable
to those that do, e.g., ParaView and VisIt.
Maintaining an API that is familiar to users while extending it to
a data-distributed scenario poses some challenges.
Furthermore, we would like to seamlessly support existing
OSPRay modules, which have added new
geometries~\cite{wald_cpu_2015,vierjahn_interactive_2017,wang_cpu_2019}
and volumes~\cite{rathke_simd_2015,wald_cpu_2017},
in a data-distributed setting.




We implement the data-distributed API through the addition
of a new OSPRay API backend, the \texttt{MPIDistributedDevice}.
As in single process rendering, each rank sets up its local geometries
and volumes independently and places them into one or more
\texttt{OSPModel} objects.
However, instead of a single model per-scene, the application
must create one model for each disjoint brick of data on the rank.
Each brick may contain any combination of geometries and
volumes, including ones provided by user modules.
To allow applications to pass OSPRay information about the data distribution,
the distributed device extends the \texttt{OSPModel} with two additional
parameters: a required integer ID, and an optional bounding box.

The ID is used to determine if two ranks have the same brick of
data and can share the rendering work using the mixed-parallel renderer.
A typical data-parallel application with a single model per-rank
could simply use the MPI rank as the ID,
while an application with a hybrid data distribution
would have a list of models and assign a unique ID for each shared brick of data.
An MPI-parallel application can even use the
distributed API for image-parallel rendering by specifying the same data
and ID on each rank.

The bounding box parameter can be used to override the model's computed
bounds, if the model contains additional ghost geometries or voxels
that should be hidden from camera rays. An additional set of
ghost models can also be passed to the renderer, containing data visible only
to secondary rays.
The bounding box parameter and ghost models allow applications to support
local shading effects such as ambient occlusion, or compute
shadows and reflections on the replicated data in the scene.


\section{Results}
\label{sec:results}
We evaluate the performance of the Distributed FrameBuffer
on the rendering algorithms described in \Cref{sec:dfb-apps},
using our implementations within OSPRay.
The benchmarks are run on two HPC systems,
the Texas Advanced Computing Center's \stampede, and
Argonne National Laboratory's \theta,
on a range of typical image- and data-parallel
rendering use cases (\Cref{fig:benchmark-datasets}).
We also perform a direct comparison
of our sort-last compositing implementation using the DFB
against IceT for a typical data-parallel use case.
To measure performance as the rendering workload varies,
the benchmarks are taken while rendering a rotation around the data set.
Unless otherwise stated, we plot the median performance for the benchmarks,
with the median absolute deviation shown as error bars. These measures
are more robust to outliers, giving some robustness 
against influence from other jobs on the system.
All benchmarks are run with one MPI rank per-node, as OSPRay uses
threads on a node for parallelism.

\stampede and \theta consist of 4200 and 4392 \INTELXEONPHI KNL
processors respectively. \stampede uses the 7250 model,
with 68 cores, while \theta uses the 7230 model with 64 cores.
\stampede contains an additional partition of 1736 dual-socket \INTELXEONPHI
Platinum 8160 SKX nodes.
Although the KNL nodes of both machines are similar, the network
interconnects differ significantly, which can effect the performance
of communication in the DFB\@. 
\stampede employs an Intel Omni-Path network in a fat-tree
topology, while \theta uses a Cray Aries network with a three-level
Dragonfly topology.



\subsection{Image-Parallel Rendering Performance}
\label{sec:results-image-parallel}
\begin{figure}
	\centering
	\vspace{-0.5em}
	\includegraphics[width=0.95\columnwidth]{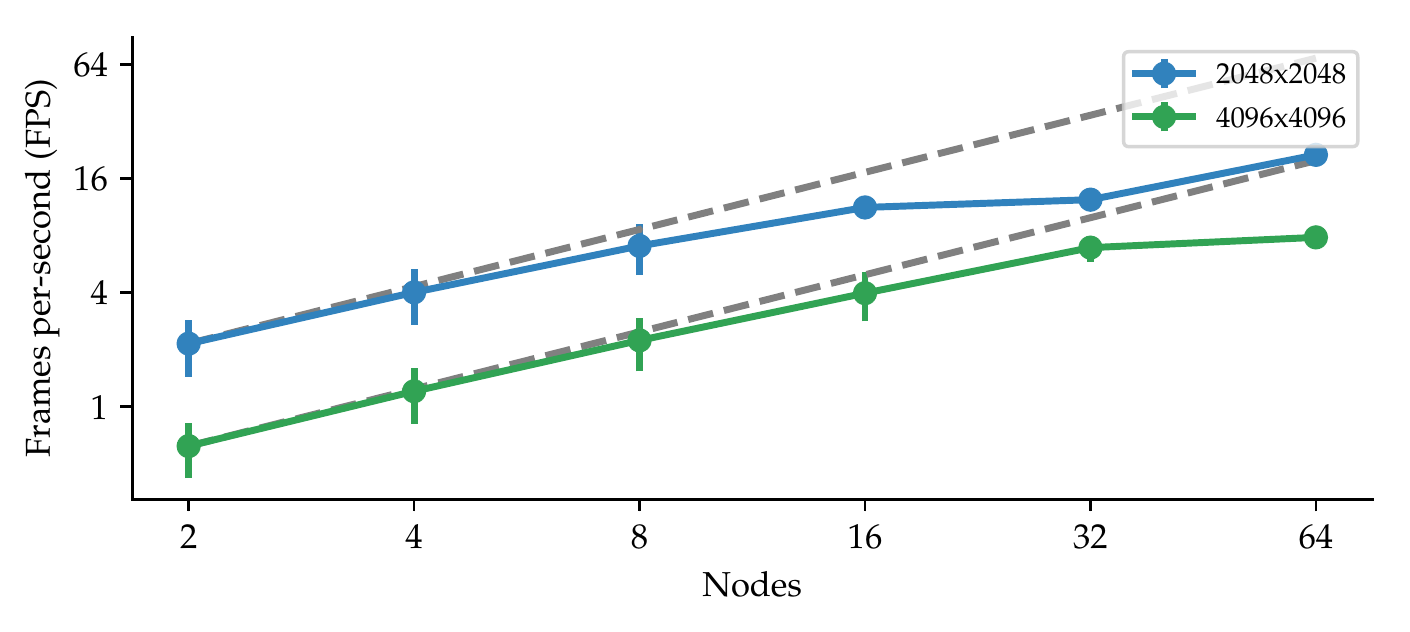}
	\vspace{-1.5em}
	\caption{\label{fig:rm-image-parallel}%
	Image-parallel strong-scaling on the R-M transparent isosurfaces
	data set on \emph{Stampede2} SKX nodes.
	The image-parallel renderer using the DFB scales
	to provide interactive rendering of expensive, high-resolution scenes.}
	\vspace{-1.5em}
\end{figure}

To study the scalability of the DFB and the image-parallel rendering
algorithm described in \Cref{sec:dfb-apps-img-parallel}, we
perform a strong scaling benchmark
using OSPRay's scientific visualization renderer. We use VTK
to extract two isosurfaces from the Richtmyer-Meshkov volume,
which are rendered with transparency and ambient
occlusion (\Cref{fig:rm-iso-benchmark}).
We measure strong-scaling on \stampede SKX nodes
at two image resolutions (\Cref{fig:rm-image-parallel}).
Although the renderer begins to drop off from the ideal scaling trend
as the local work per-node decreases, this could potentially
be addressed by employing the work-subdivision and load-balancing
strategies discussed in \Cref{sec:dfb-apps-load-balancing}.


\subsection{Data-Parallel Rendering Performance}
\label{sec:results-data-parallel}
To study the scalability of the DFB when applied to the standard data-parallel
rendering algorithm in \Cref{sec:dfb-apps-data-parallel},
we run strong scaling
benchmarks with two large-scale data sets on \stampede
and \theta.  On \stampede we render a combined visualization
of the DNS with transparent isosurfaces (\Cref{fig:dns-iso-vol-benchmark}),
and on \theta we render the $5^3$ Cosmic Web subset (\Cref{fig:cosmic-web-benchmark}).
We find that our data-parallel renderer using the DFB is able to provide 
interactive frame rates for these challenging scenes, and scale up
performance with more compute.

\begin{figure}
	\centering
	\includegraphics[width=0.95\columnwidth]{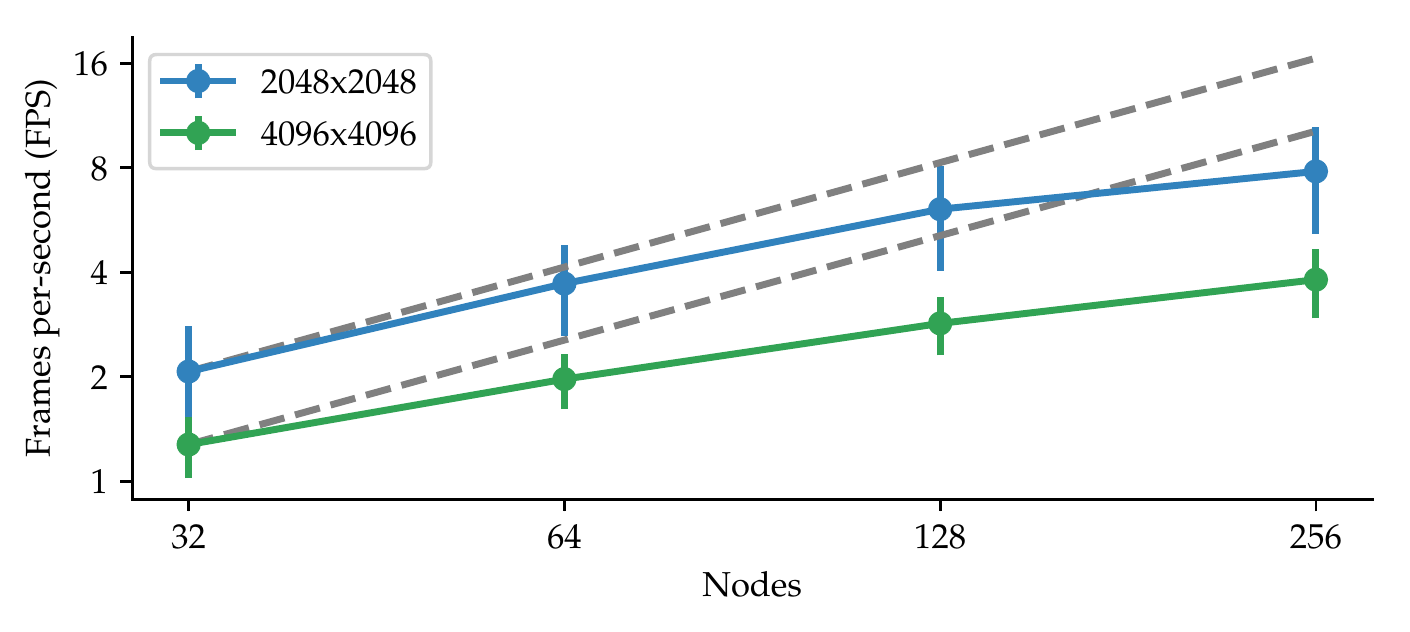}
	\vspace{-1.5em}
	\caption{\label{fig:cosmic-web-data-parallel}%
	Data-parallel strong-scaling on the Cosmic Web data set on \theta. We find
	close to ideal scaling at moderate image sizes and node counts, with
	somewhat poorer scaling at very high resolutions.}
	\vspace{-1em}
\end{figure}

On the Cosmic Web we observe good scaling from 32 to 64 nodes
(\Cref{fig:cosmic-web-data-parallel}). Although performance begins to trail
off the ideal trend beyond 128 nodes, absolute rendering performance remains
interactive.

\begin{figure}
	\centering
	\begin{subfigure}{\columnwidth}
		\centering
		\includegraphics[width=0.95\textwidth]{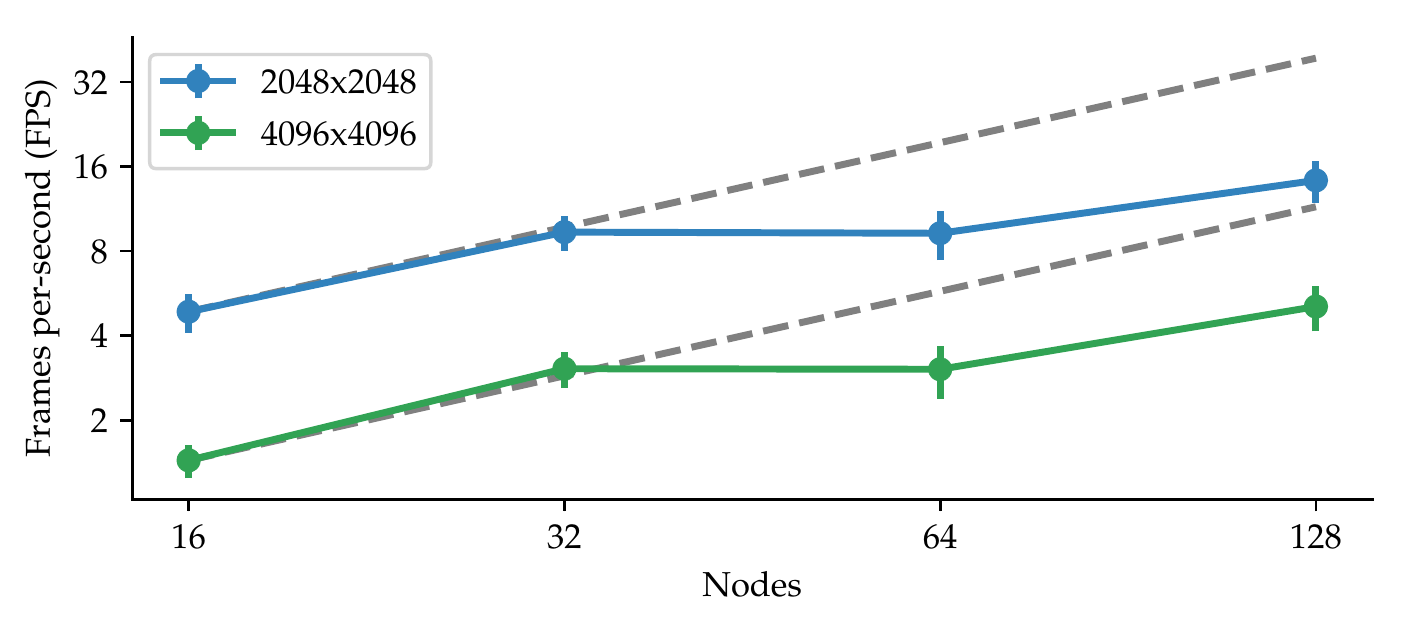}
		\vspace{-1em}
		\caption{\label{fig:dns-iso-vol-data-parallel-overall}%
		Overall rendering performance.}
	\end{subfigure}
	\begin{subfigure}{\columnwidth}
		\centering
		\includegraphics[width=0.95\textwidth]{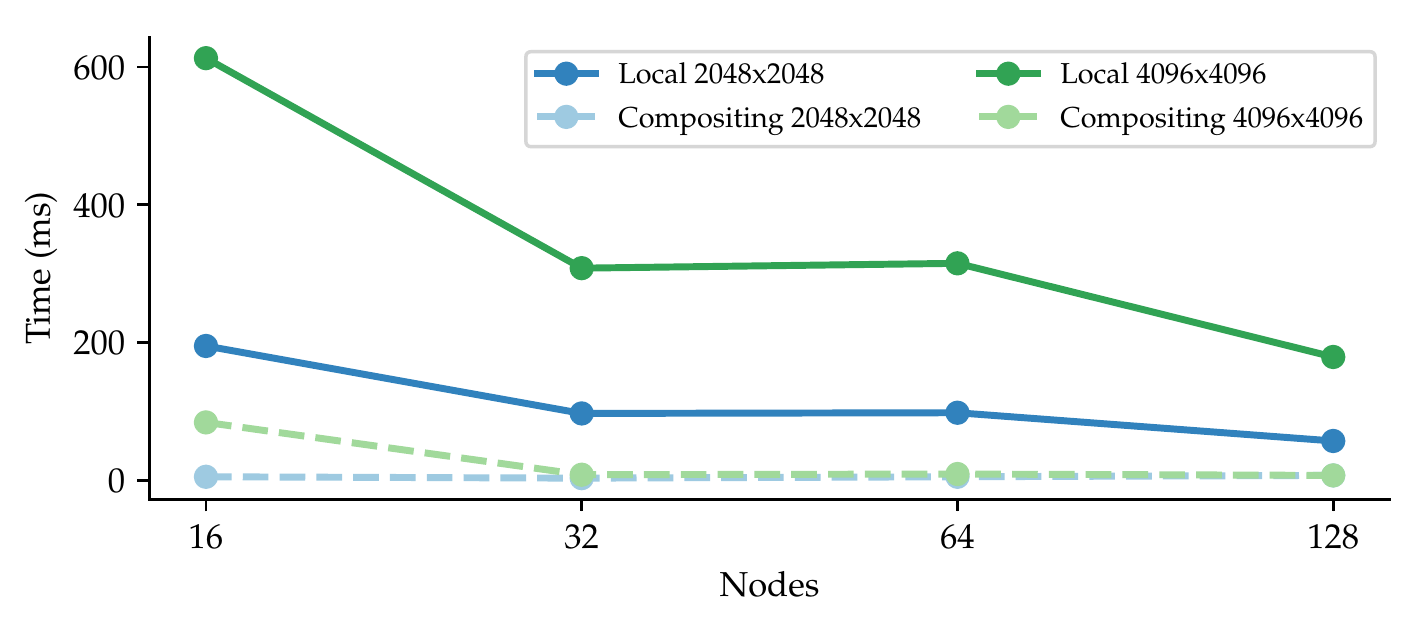}
		\vspace{-1em}
		\caption{\label{fig:dns-iso-vol-data-parallel-breakdown}%
		Timing breakdown of local rendering and compositing overhead.}
	\end{subfigure}
	\vspace{-1em}
	\caption{\label{fig:dns-iso-vol-data-parallel}%
	Data-parallel strong-scaling on the DNS with isosurfaces on \stampede KNLs.
	The lack of scaling from 32 to 64 nodes is attributable to a poor local
	work distribution (b), which can be partially addressed by using our
	mixed-parallel renderer.}
	\vspace{-1.5em}
\end{figure}

On the DNS we
find near ideal scaling from 16 to 32 nodes (\Cref{fig:dns-iso-vol-data-parallel-overall});
however, we observe little change from 32 to 64 nodes, although we
see improvement again at 64 to 128 nodes.
To find the cause of the bottleneck at 64 nodes,
we look at a breakdown of the
time spent rendering the rank's local data and the
compositing overhead incurred by the
DFB (\Cref{fig:dns-iso-vol-data-parallel-breakdown}).
Compositing overhead refers to the additional time the compositor takes to complete the
image, after the slowest local rendering task has completed~\cite{grosset_dynamically_2016}.
In this case we find that the bottleneck is caused by the local
rendering task not scaling, which could be addressed by employing a
hybrid data distribution or the work-splitting techniques discussed previously.

\subsubsection{Compositing Performance Comparison with IceT}
\label{sec:results-compare-icet}
\begin{figure*}
	\vspace{-0.5em}
	\centering
	\begin{subfigure}{0.33\textwidth}
		\centering
		\includegraphics[width=\textwidth]{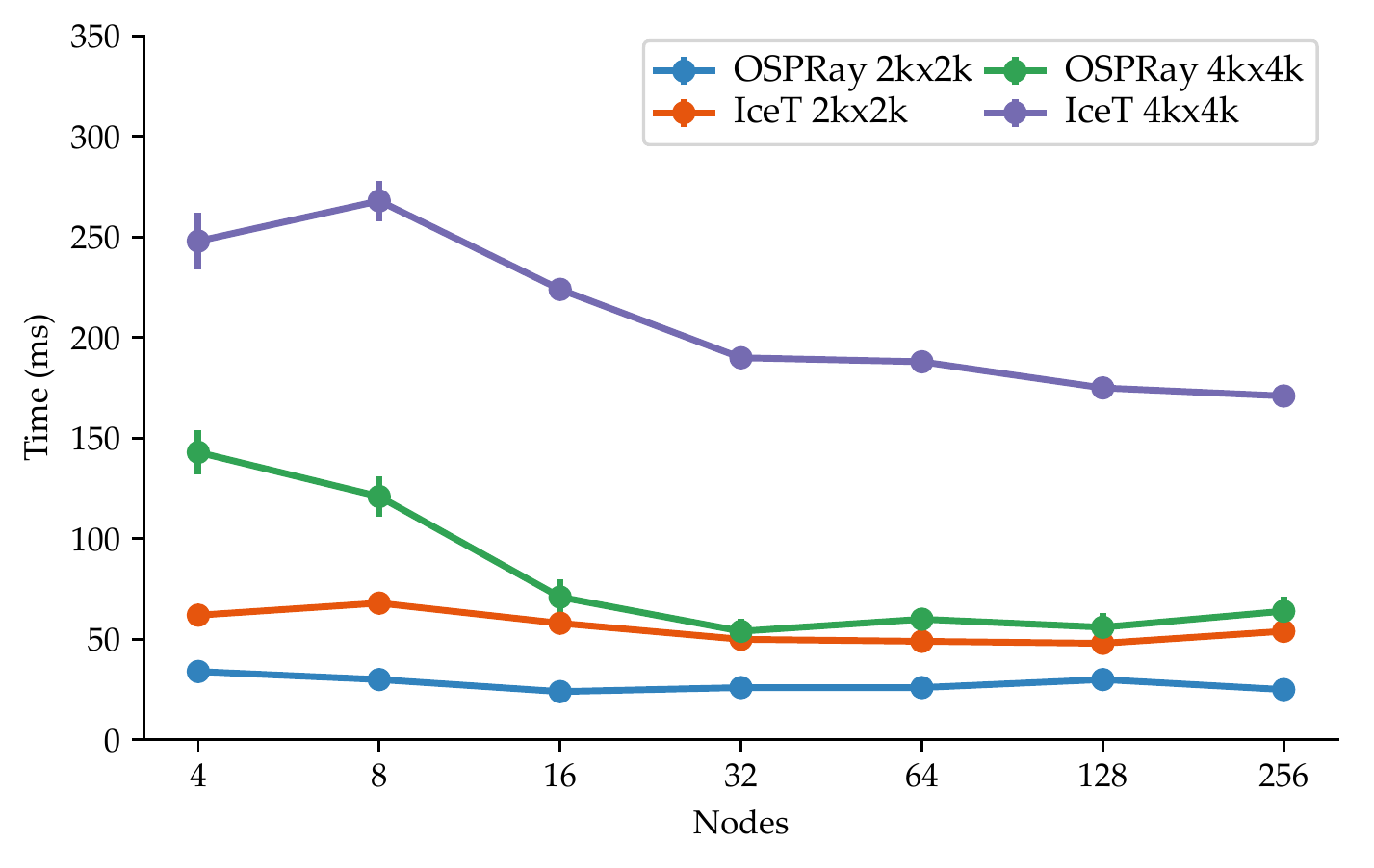}
		\vspace{-2em}
		\caption{\label{fig:icet-overall-theta-knl}%
		\theta total time.}
	\end{subfigure}%
	\begin{subfigure}{0.33\textwidth}
		\centering
		\includegraphics[width=\textwidth]{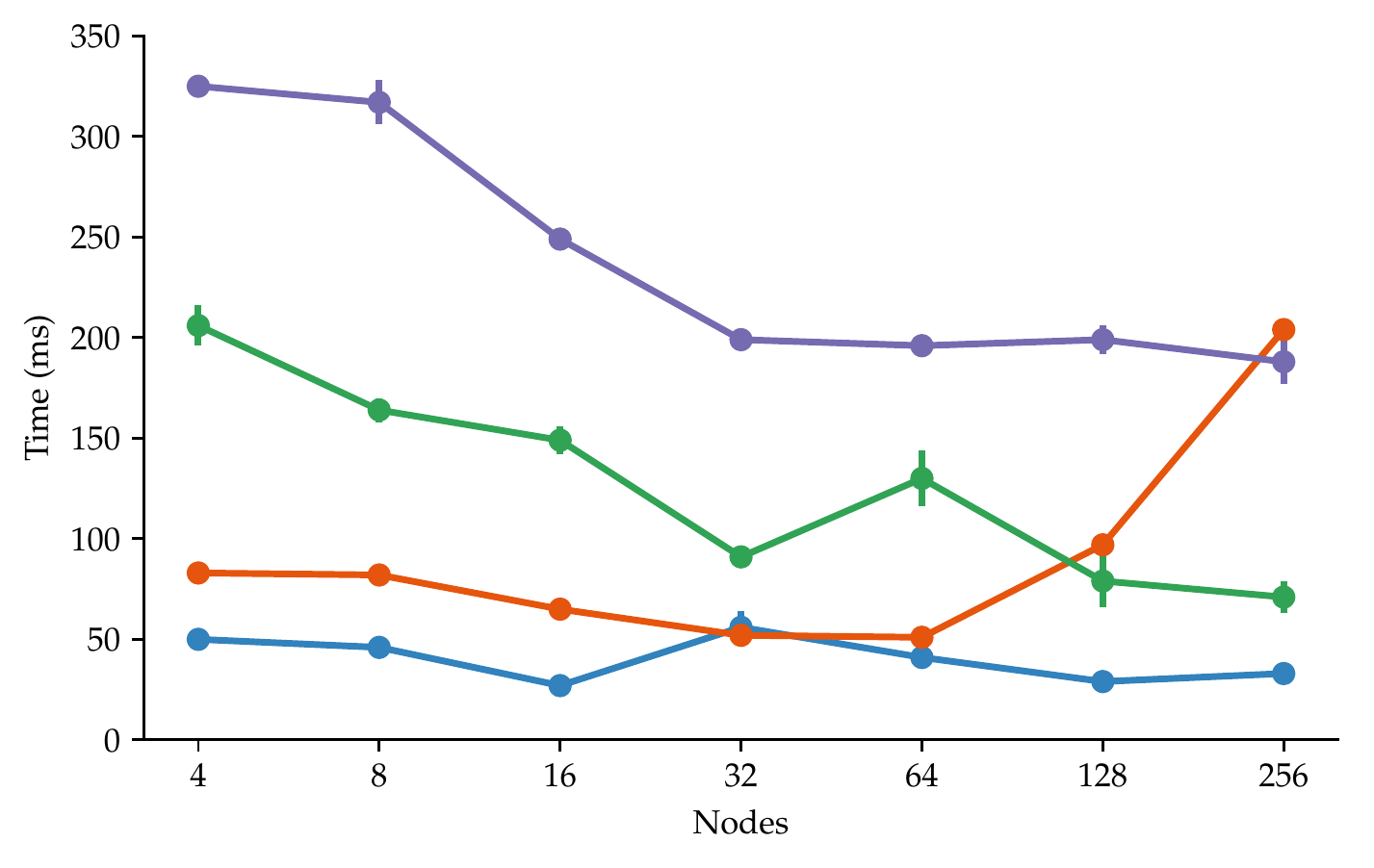}
		\vspace{-2em}
		\caption{\label{fig:icet-overall-stamp2-knl}%
		\stampede KNL total time.}
	\end{subfigure}%
	\begin{subfigure}{0.33\textwidth}
		\centering
		\includegraphics[width=\textwidth]{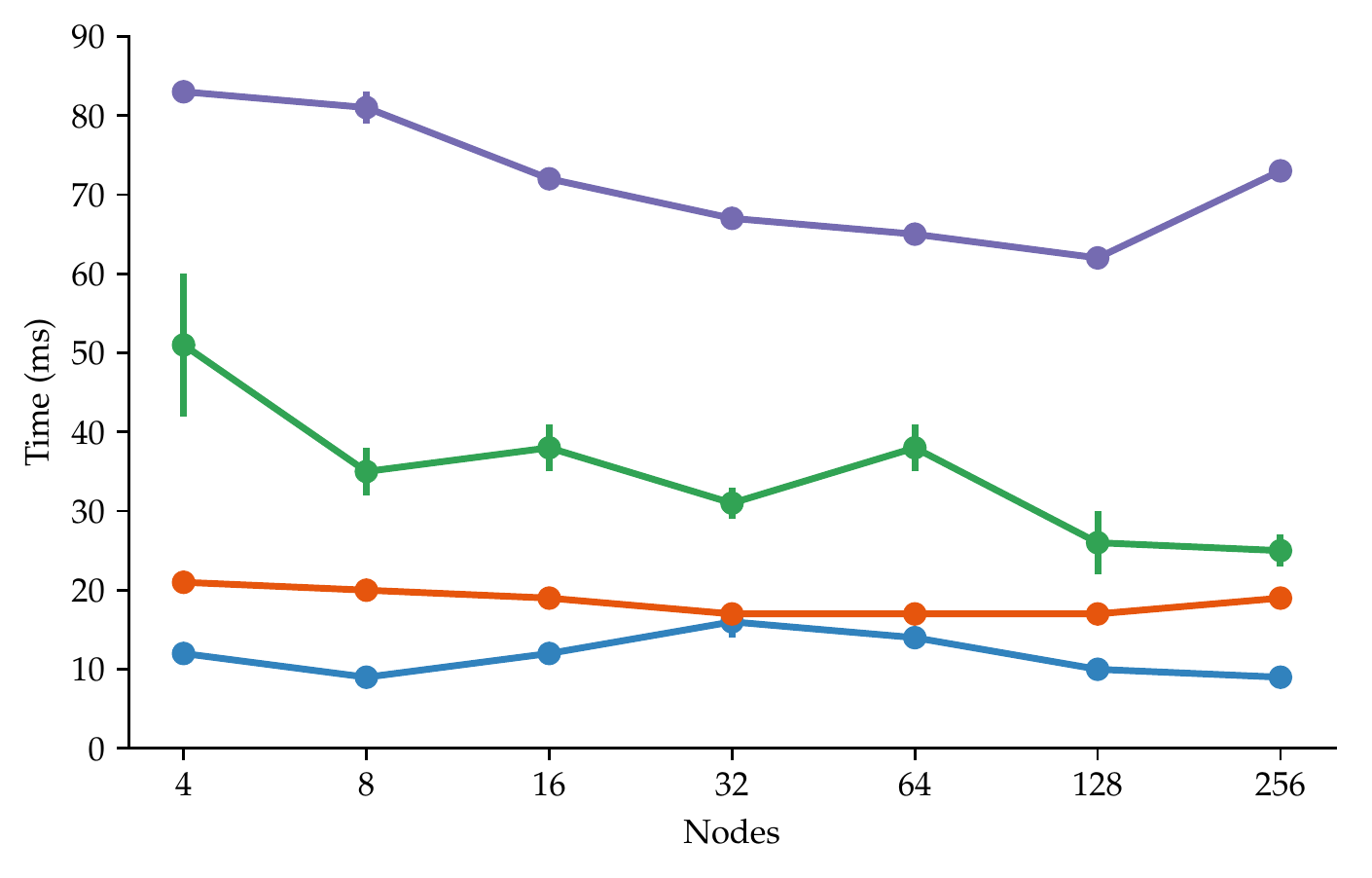}
		\vspace{-2em}
		\caption{\label{fig:icet-overall-stamp2-skx}%
		\stampede SKX total time.}
	\end{subfigure}
	\begin{subfigure}{0.33\textwidth}
		\centering
		\includegraphics[width=\textwidth]{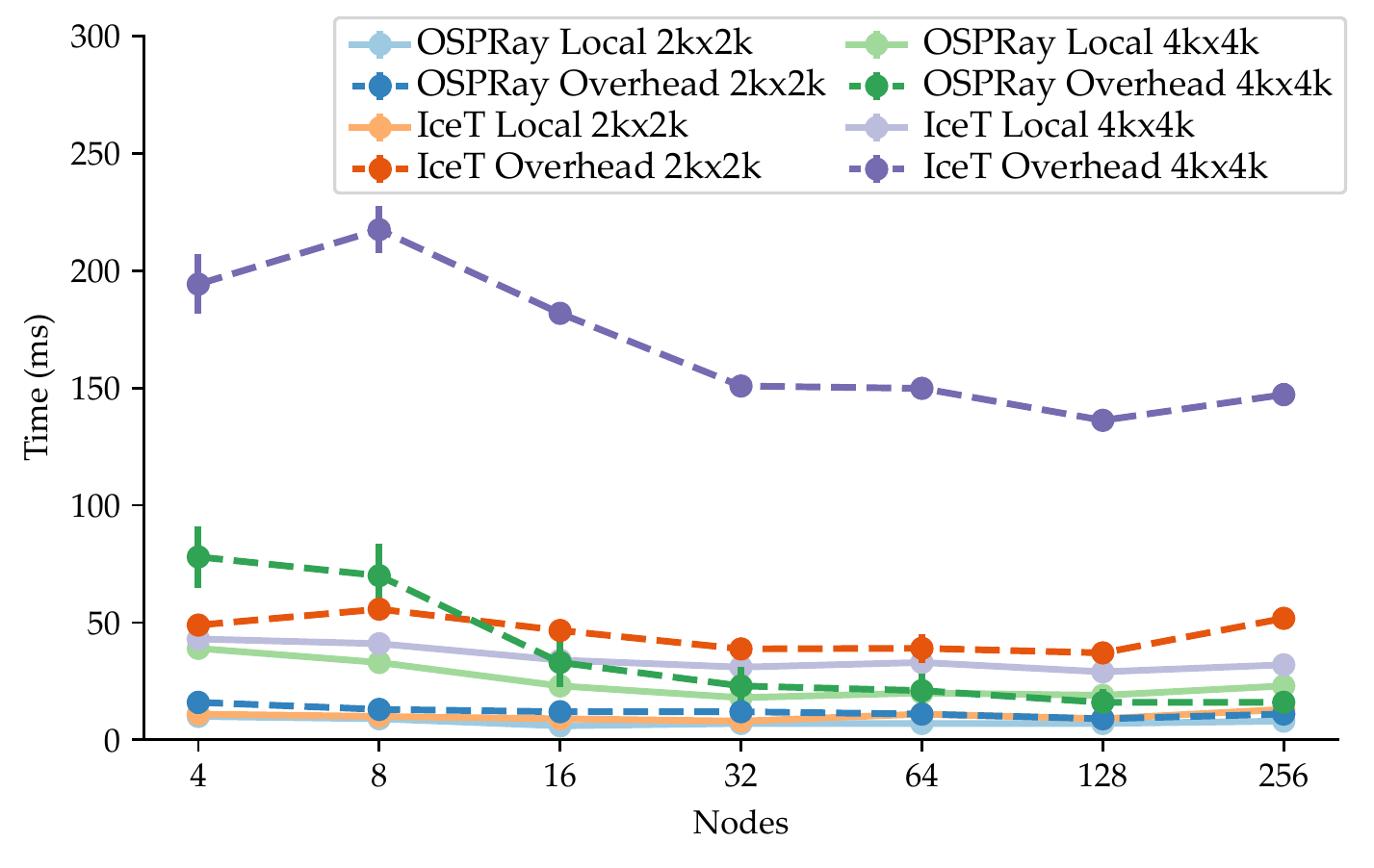}
		\vspace{-2em}
		\caption{\label{fig:icet-breakdown-theta-knl}%
		\theta timing breakdown.}
	\end{subfigure}%
	\begin{subfigure}{0.33\textwidth}
		\centering
		\includegraphics[width=\textwidth]{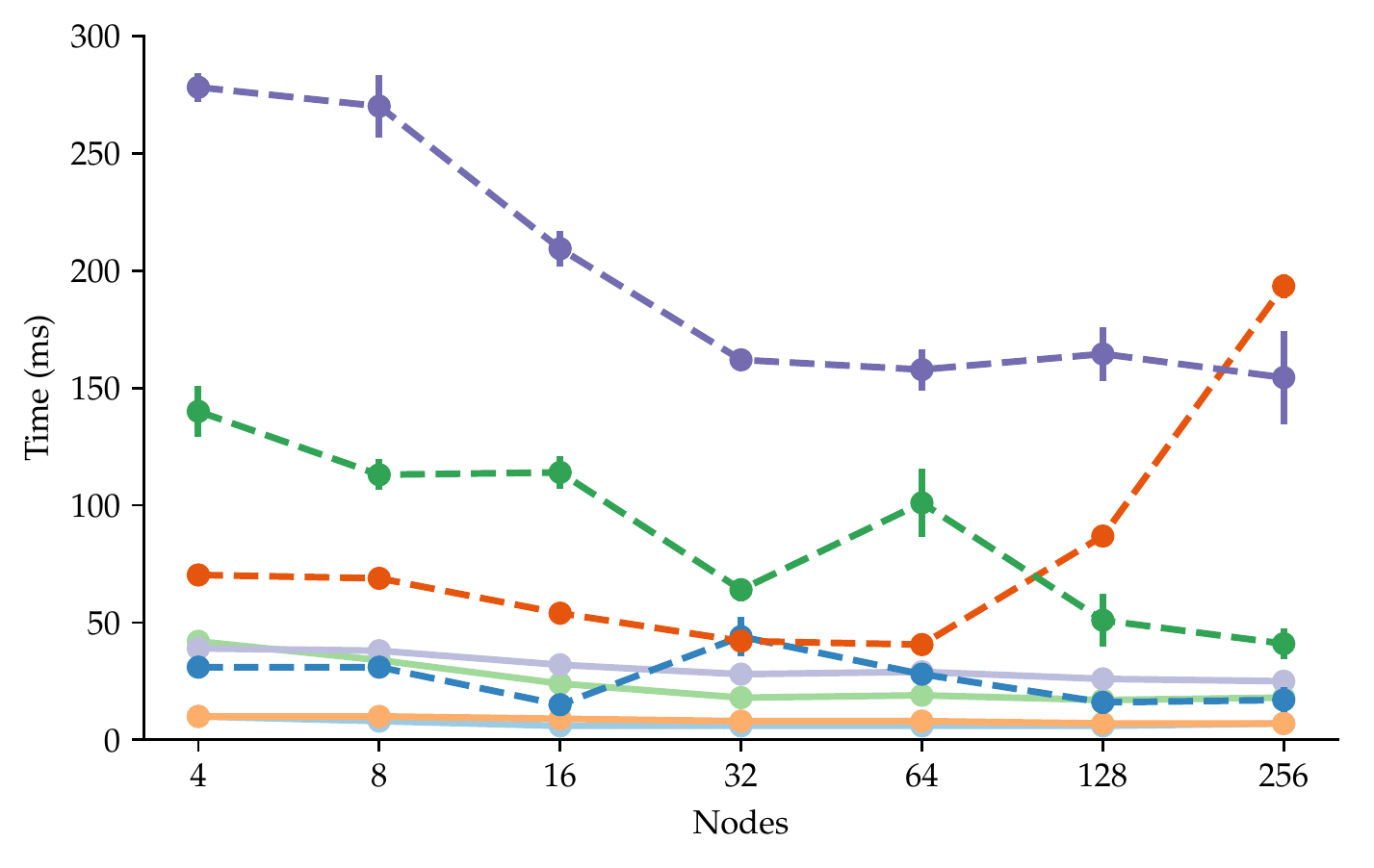}
		\vspace{-2em}
		\caption{\label{fig:icet-breakdown-stamp2-knl}%
		\stampede KNL timing breakdown.}
	\end{subfigure}%
	\begin{subfigure}{0.33\textwidth}
		\centering
		\includegraphics[width=\textwidth]{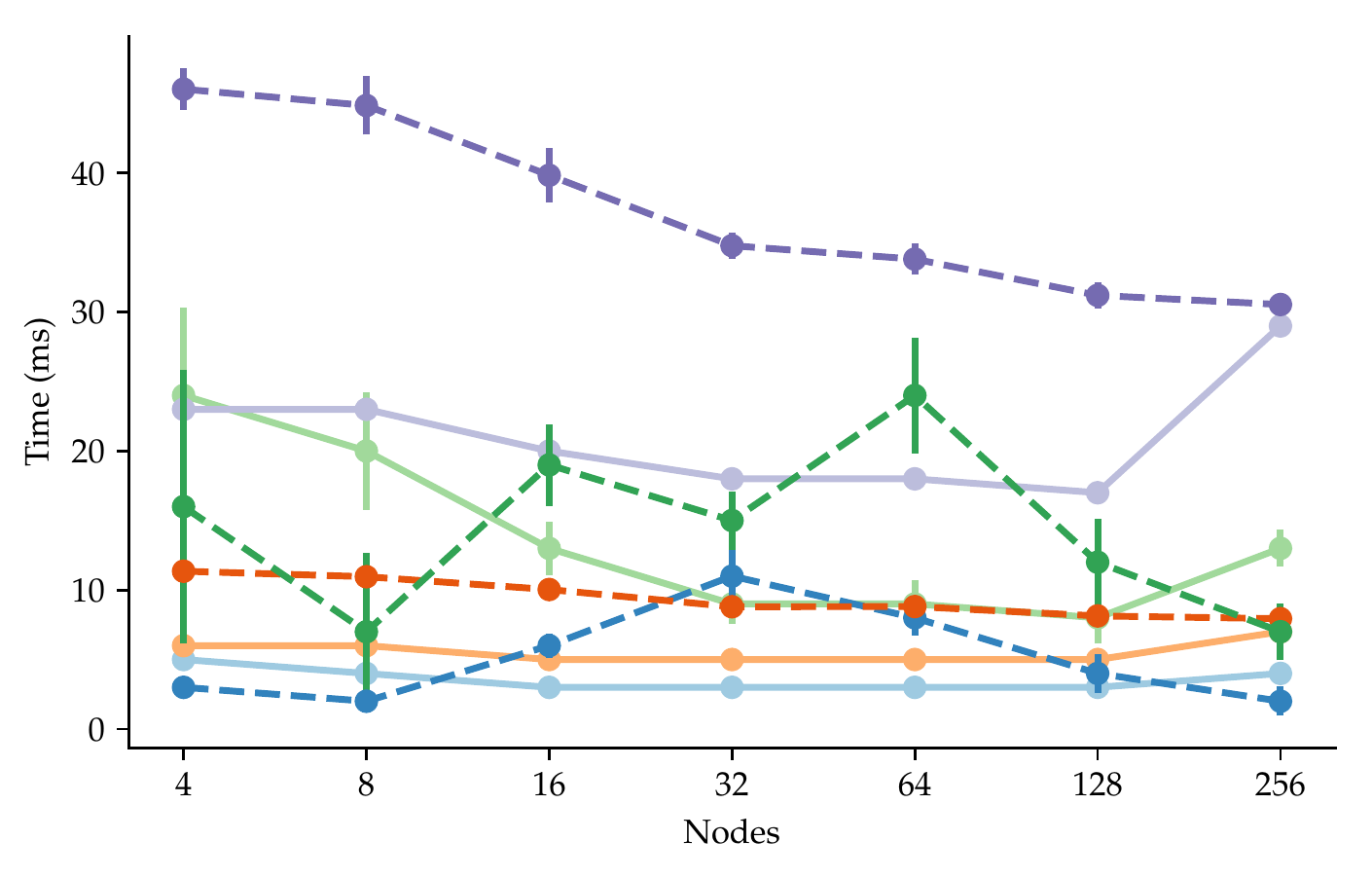}
		\vspace{-2em}
		\caption{\label{fig:icet-breakdown-stamp2-skx}%
		\stampede SKX timing breakdown.}
	\end{subfigure}
	\vspace{-1em}
	\caption{\label{fig:icet-compare}%
	Compositing benchmark performance comparison of the DFB and IceT on the synthetic data set.
	We find that our approach achieves better, or at least similar, scaling as IceT,
	while providing faster absolute rendering times. In the timing breakdowns (d-f),
	we observe this difference is due to the DFB achieving a significant
	reduction in compositing overhead.}
	\vspace{-1.5em}
\end{figure*}

To perform a direct comparison with IceT for data-parallel
rendering, we use a synthetic data set (\Cref{fig:compositing-benchmark}),
and modify our data-parallel renderer to support using IceT for compositing.
The IceT renderer follows the same code-path as our data-parallel renderer
to render its assigned brick of data, then
hands the framebuffer off to IceT for compositing.
We found IceT's automatic compositing algorithm selection
to give the best performance, and use this mode throughout the benchmarks.

In terms of overall scalability and performance, our approach
scales better then, or at least similar to, IceT, while achieving better absolute
rendering performance (\Cref{fig:icet-overall-stamp2-knl,fig:icet-overall-theta-knl,fig:icet-overall-stamp2-skx}).
When comparing timing breakdowns (\Cref{fig:icet-breakdown-stamp2-knl,fig:icet-breakdown-theta-knl,fig:icet-breakdown-stamp2-skx})
we find that, as expected, local rendering times are similar, and
the performance difference is due to the
differing compositing overhead.
It is important to note that some of the absolute
difference in overhead is due to IceT's synchronous design,
which makes it unable to overlap compositing with rendering.
We can consider a hypothetical IceT implementation which does overlap
compositing and rendering by subtracting the local rendering time from the compositing
overhead, and find that the DFB still achieves similar or superior compositing performance.
Furthermore, we observe that when comparing the scaling trends
of the two approaches, the DFB scales similar to, or better than, IceT.
Although a rigorous comparison is difficult due to the different
HPC systems used, the DFB follows similar scaling trends as Grosset et al.'s DSRB~\cite{grosset_dynamically_2016},
while providing greater flexibility.

Finally, we evaluate the portability of our approach
by comparing the KNL runs on \stampede (\Cref{fig:icet-overall-stamp2-knl,fig:icet-breakdown-stamp2-knl})
and \theta (\Cref{fig:icet-overall-theta-knl,fig:icet-breakdown-theta-knl}).
The slightly different KNLs on each system
will have a minor effect on performance; however
any significant differences are
attributable to the differing network architectures and
job placement strategies.
On \stampede we observe a rather bumpy scaling trend where,
depending on the image size, we see a temporary decrease in the compositing
performance at certain node counts. On \theta we observe
a smoother trend, with better absolute compositing performance.
We found that disabling message compression on \theta gave
better performance, while on \stampede we encountered
MPI messaging performance issues at 16 nodes and up without it.
Thus, we leave
compression as an option to users which is enabled by default
at 16 nodes. In our benchmarks we disable compression on \theta,
and enable it at 16 nodes and up on \stampede.
IceT uses a custom image compression
method, which is not easily disabled.


\subsection{Hybrid Data Distribution Rendering Performance}
\label{sec:results-hybrid-parallel}
\label{sec:results-mixed-parallel}
\begin{figure}
	\centering
	\begin{subfigure}{0.48\textwidth}
		\centering
		\includegraphics[width=\textwidth]{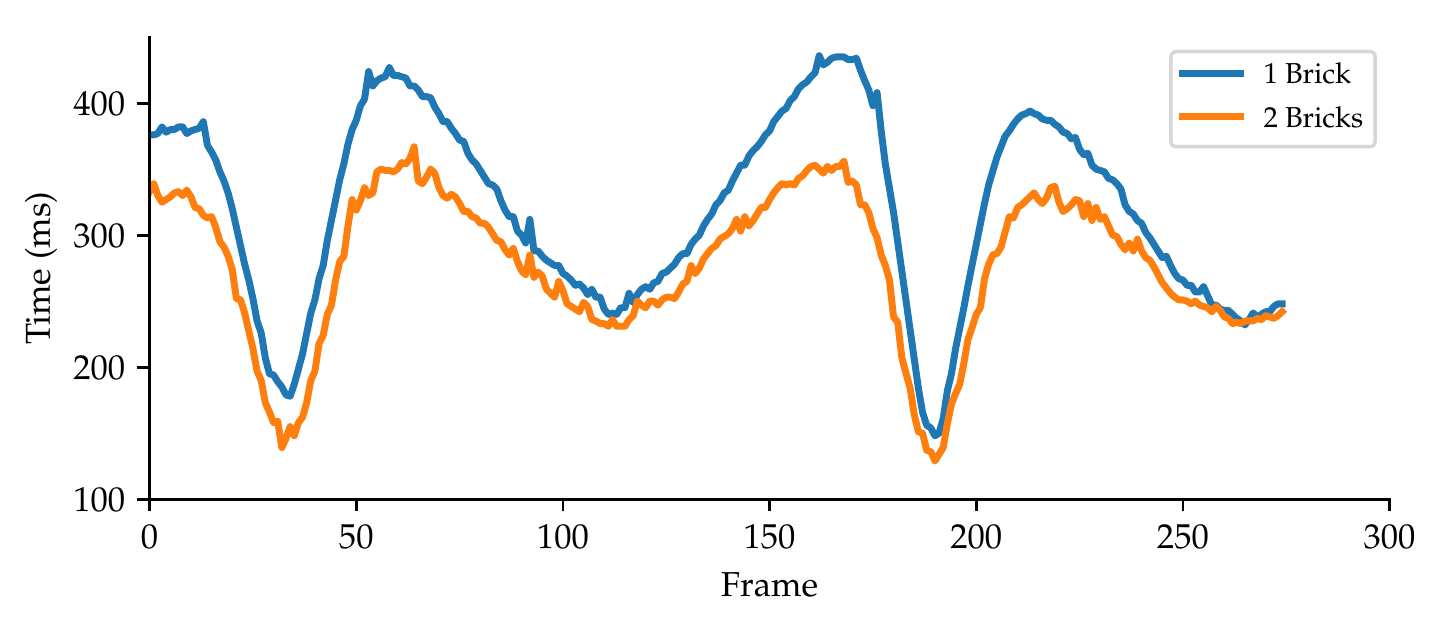}
		\vspace{-2em}
		\caption{\label{fig:hybrid-dns-64}%
		Per-frame render time on 64 \stampede KNLs, at $4096\times4096$} 
	\end{subfigure}
	\begin{subfigure}{0.48\textwidth}
		\centering
		\includegraphics[width=\textwidth]{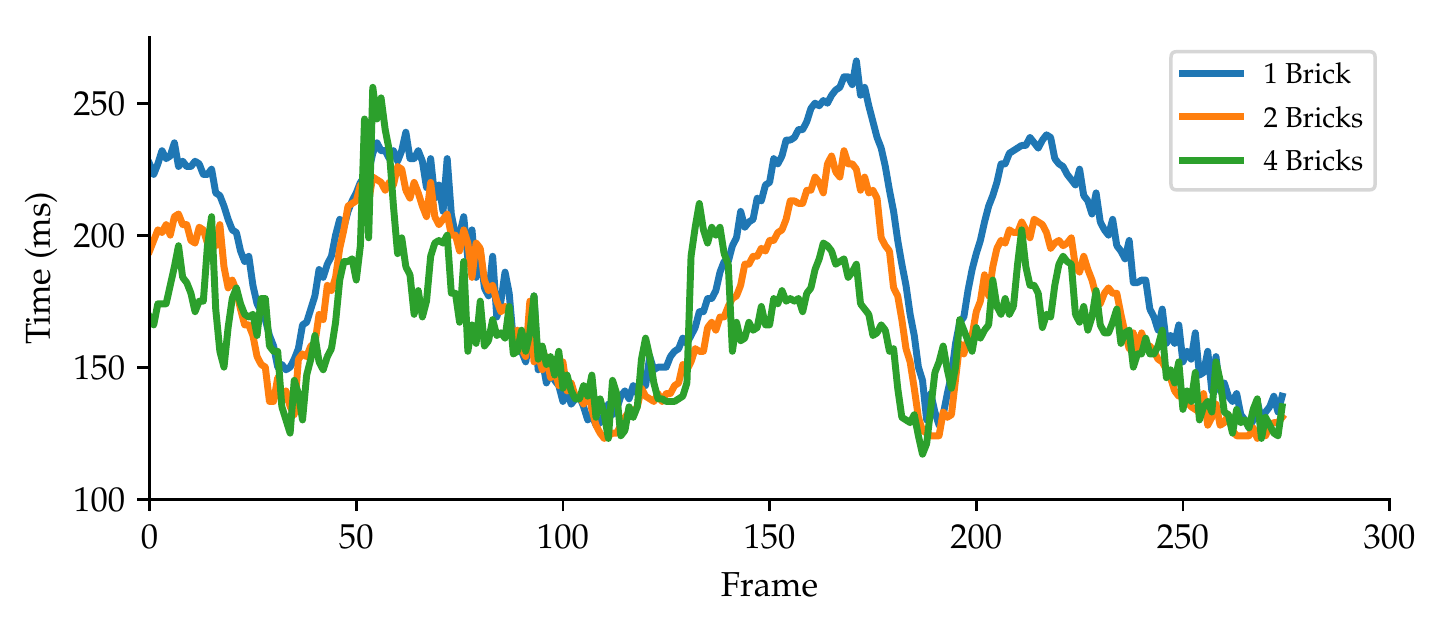}
		\vspace{-2em}
		\caption{\label{fig:hybrid-dns-128}%
		Per-frame render time on 128 \stampede KNLs, at $4096\times4096$} 
	\end{subfigure}
	\vspace{-1em}
	\caption{\label{fig:hybrid-render-dns}%
	Improving load-balancing on the DNS with isosurfaces with partial data-replication
	in the mixed-parallel renderer.
	Sharing rendering between two nodes (two bricks per-node) gives a consistent improvement,
	between four tends to give further improvement.}
	\vspace{-2em}
\end{figure}

To measure the impact of partial data replication on
load balance, we look at the per-frame overall time on the DNS
with isosurfaces data set on \stampede (\Cref{fig:hybrid-render-dns}).
The volume is partitioned into as many bricks as there are ranks,
with bricks redundantly assigned to ranks based on the available memory capacity.
When using 64 KNLs there is enough memory to store two bricks per-rank,
with 128 KNLs we can store up to four. The rendering work for each brick will be
distributed among two or four ranks, respectively.
The redundant bricks are distributed using a simple round-robin assignment.
A brick distribution based on, e.g., some space filling curve or runtime tuning,
could provide additional improvement.

In both the 64 and 128 node runs the two brick per-node configuration provides
a consistent improvement over no replication.
This improvement is more pronounced for camera positions with greater load
imbalance.
With four bricks per-node, there are larger fluctuations
in rendering performance, though at times we do find improvement
over the two brick configuration.
These larger fluctuations could be due
to increased memory traffic,
which is alleviated as data is cached in the KNL MCDRAM.
This theory is further supported by the sharp spikes in performance,
when new data must be fetched from RAM.

\section{Conclusion}
\label{sec:conclusion}
We have presented the Distributed FrameBuffer,
an asynchronous, distributed image processing and compositing framework
primarily targeted at rendering applications.
By breaking the image processing operations into a set of per-tile
tasks with independent dependency trees, the DFB simplifies
the implementation of complex distributed rendering algorithms.
Moreover, the DFB does not trade performance for this flexibility
and we report performance competitive with specialized state-of-the-art
algorithms. Our data-distributed API extension to OSPRay
has already been used successfully in practice
for in situ visualization~\cite{usher_libis_2018}.

We have merged our implementation of the DFB, the
rendering algorithms presented, and the data-distributed API into OSPRay,
and released them in version 1.8.
While prior work integrated OSPRay into VisIt~\cite{wu_visit-ospray_2018}
by using OSPRay's single-node rendering API and IceT for compositing,
this can now be done using the distributed API directly.
Compared to
results reported on prior versions of OSPRay~\cite{abram_galaxy_2018}
our work provides significant performance improvements.

However, the DFB and rendering algorithms presented are not
without limitations. The rendering algorithms presented
support only local lighting effects computed with the data available
on a rank.
Although approaches to compute global illumination on distributed data
by sending rays between nodes~\cite{abram_galaxy_2018,park_spray_2018}
could be implemented in the DFB, it is unclear how well
a naive implementation would perform, or if extensions
to the DFB would be required. We leave this exciting
avenue of research as future work.

In our evaluation we observed large differences
in MPI performance and network behavior between \stampede and \theta.
Although we expose the use of compression as an option
for users to tune as needed, it would be worthwhile
to investigate self-tuning strategies for the DFB to automatically
adapt to such architectural differences.


\section*{Acknowledgments}
We would like to thank Damon McDougall and Paul Navr\'atil of
the Texas Advanced Computing Center
for assistance investigating MPI performance at TACC, and
Mengjiao Han for help with the display wall example.
The Cosmic Web and DNS datasets were made available by Paul
Navr\'atil, the Richtmyer-Meshkov is courtesy of Lawrence Livermore National Laboratory.
This work is supported in part by the Intel Parallel Computing Centers Program,
NSF: CGV Award: 1314896, NSF:IIP Award: 1602127,
NSF:ACI Award: 1649923, DOE/SciDAC DESC0007446, CCMSC DE-NA0002375 and NSF:OAC Award: 1842042.
This work used resources of the Argonne Leadership Computing Facility,
which is a U.S. Department of Energy Office of Science User Facility supported under
Contract DE-AC02-06CH11357. 
The authors acknowledge the Texas Advanced Computing Center (TACC) at The University of Texas
at Austin for providing HPC resources that have contributed to the research results reported
in this paper.

\bibliographystyle{eg-alpha-doi}
\bibliography{bibliography}

\end{document}